\documentclass[12pt,preprint]{aastex}

\shorttitle{Spectra of Young Nearby Stars}
\shortauthors{White et al. 2006}

\usepackage{amsmath}
\usepackage{natbib}

\begin{document}

\title{High Dispersion Optical Spectra of Nearby Stars Younger Than the Sun}

\author{Russel J. White\altaffilmark{1}, Jared M. Gabor\altaffilmark{2}, Lynne A. Hillenbrand}
\affil{Department of Astronomy, California Institute of Technology, MS
  105-24, Pasadena, CA 91125}
\email{russel.white@uah.edu, jgabor@orion.as.arizona.edu, lah@warning.caltech.edu}

\altaffiltext{1}{Present Address: University of Alabama in Huntsville, 301 Sparkman Dr., OB 201 B, 
Huntsville, AL 35899}

\altaffiltext{2}{Present Address: University of Arizona, 933 N. Cherry Ave, Rm. N204, Tucson, AZ
85721}

\begin{abstract}
We present high dispersion (R $\sim$ 16,000) optical ($3900 - 8700$ \AA)
spectra of 390 stars obtained with 
the Palomar 60-inch Telescope.  The majority of stars
observed are part of the Spitzer Legacy Science Program \textit{The
Formation \& Evolution of Planetary Systems}.  Through detailed analysis
we determine stellar properties for this sample, including radial and rotational
velocities, Li\,I $\lambda$ 6708 and $H\alpha$ equivalent widths, the chromospheric
activity index $R'_{HK}$, and temperature- and gravity-sensitive 
line ratios.  Several spectroscopic binaries are also identified.  From our tabulations, 
we illustrate basic age- and rotation-related correlations among measured indices.
One novel result is that Ca\,II chromospheric emission
appears to saturate at $v$sin$i$ values above $\sim 30$ km s$^{-1}$, similar to the well
established saturation of x-rays that originate in the spatially separate coronal 
region.
\end{abstract}

\keywords{stars: fundamental parameters --- stars: activity --- stars: pre-main sequence}

\section{Introduction}

We have obtained high dispersion spectra of several hundred solar type stars
within 20-160 pc of the Sun in order to determine fundamental stellar properties.
Our work is in support of the {\it Spitzer}/Legacy Program ``Formation and 
Evolution of Planetary Systems" \citep[FEPS;][]{meyer06}, which
aims to understand the evolution of circumstellar dust from 
the primordial planet-building phase at an age of $\sim 3$ Myr to mature debris disk
systems up to an age of $\sim 3$ Gyr.  While data from the {\it Spitzer} Space
Telescope probe the dust characteristics, a wide range of ancillary 
ground-based observations are required to estimate stellar properties. 
High dispersion spectroscopy, in particular, is a valuable tool for 
determining radial and rotational velocities, effective temperatures and 
surface gravities, abundances, and chromospheric activity diagnostics.
These in turn provide useful constraints on stellar masses and ages.
The stellar parameters derived from these spectroscopic observations 
thereby permit multiple axes of investigation for the infrared data from
{\it Spitzer}, such as whether dust disk characteristics can be
(anti-)correlated with stellar mass, signatures of youth, and/or stellar 
multiplicity.

In Section 2 we describe the observed sample and the spectroscopic observations.  
In Sections 3 and 4 we summarize the spectroscopic reductions and the methods 
used to extract spectroscopic properties.  In Section 5 we discuss the spectroscopic 
binaries identified in the sample and illustrate basic age related correlations among 
the measured properties; the relation and apparent saturation of chromospheric Ca\,II 
emission versus projected rotational velocity is highlighted.
The analysis is meant to be representative rather than complete.  The 
spectroscopic properties are presented primarily to assist more detailed 
studies of individual FEPS stars.

\section{Sample and Observations}

Our primary sample for high dispersion spectroscopic observations consists
of actual FEPS targets as well as many candidate targets eventually dropped from the
FEPS program.  The source list for FEPS is comprised of young near-solar 
analogs that range in mass from 0.7-1.5 M$_\odot$ and span ages 
between 3 Myr and 3 Gyr.  This source list was drawn from 
three samples of young solar-analogs.  
The first sample was assembled from the D. R. Soderblom (2000, private communication) 
volume-limited ($<$ 50 pc) spectroscopic study of stars with B-V colors between 0.52 and 0.81 
mag (spectral types ~ F8-K0) and Mv magnitudes within 1.0 mag of the solar-metallicity 
zero-age main sequence; approximate ages are provided from the chromospheric activity 
index $R'_{HK}$.
Since this nearby field sample primarily consists of stars older than a few\,$\times\, 
10^8$ yr, a second sample was assembled with an targeted age range of 3-300 Myr.
Stars in this sample were identified based on observational investigations to identify 
young main-sequence and pre-main sequence solar-type stars with strong x-ray
emission, high lithium abundances, and kinematics appropriate for the young
galactic disk \citep[e.g.][]{mamajek02}.
The third sample consists of stars in nearby well-studied clusters, including 4
open clusters [IC 2602 (55 Myr), Alpha Per (90 Myr), Pleiades (125 Myr), 
Hyades (650 Myr)] and 3 younger T Tauri age clusters [R Corona Australis (3 Myr), 
Upper Scorpius (5 Myr), Upper Centaurus Lupus (10 Myr)].  Stars in these clusters
serve to benchmark our field star selection criteria
by providing sub-samples nearly identical in age, composition, and birth
environment.  These three parent samples and the procedure for downselecting
to the final FEPS sample observed with {\it Spitzer}
are described more fully in \citet{meyer06}.  In addition to stars in this primary 
sample, a handful of other young and/or chromospherically active stars (e.g. 
RS CVn) were observed for comparison, and $\sim 20$ stars with precisely
known radial velocities \citep{nidever02} were observed for calibration and analysis
purposes.

High dispersion spectroscopic observations were obtained for 363 stars 
among the sample described above; 284 of these stars are among the 326 stars 
that comprise the final FEPS program as executed with {\it Spitzer} 
(the vast majority of FEPS stars not observed are in the southern hemisphere
and thus inaccessible from Palomar Observatory).  Fifty-two of the 363 stars
observed are members of the 3 open clusters and 31 stars are members of the
3 young clusters.  For calibration purposes,
we also observed 13 giant stars with spectral types ranging from G5 
to M2, 5 early- to mid-M dwarf stars, and 9 early A type stars with nearly
featureless continua.  Tables 1\footnote{The complete version of Table 1
is available online.} and 2 list the observed sample (excluding 
non-solar type calibration stars), ordered by right ascension; Table
1 contains spectroscopic single stars and candidate single-lined
spectroscopic binaries (for which stellar properties could be determined)
while Table 2 contains known or newly discovered double-lined spectroscopic
binaries.  Coordinates for all stars are from the 2MASS All-Sky Catalog of
Point Sources \citep{cutri03}, as interpreted by VizieR \citep{ochsenbein00}.

Spectroscopic observations were obtained on 9 observing runs 
between July 2001 and June 2003 with the Palomar 60-inch telescope 
and facility spectrograph \citep{mccarthy88} in its echelle mode.  
This instrument is no longer available.
A 1\farcs43$\times$7\farcs36 slit was used, yielding spectra spanning
from approximately 3900 \AA\, to 8700 \AA, with some small gaps between 
the redder orders.  While the 2-pixel resolving power of this instrument is
$\sim 19,000$, the achieved resolving power was typically $\sim 16,000$
because of image quality issues.
The wavelength coverage 
was chosen so as to include both the Ca\,II H\&K features at
the blue end and the Ca\,II infrared-triplet at the red end.  The spectrograph
images the echelle spectra onto an 800$\times$800 CCD; the consequence
of projecting the broad wavelength coverage onto this modest sized CCD
is that the orders are closely spaced, especially at the red end.
The detector has a gain of 1.5 and
read noise $\sim 12$ electrons.  During each night of
observations, series of bias exposures and flat field exposures were
obtained to effect standard image processing.  Because of the large 
spectral range of the
instrument, the flat field images were taken with two separate incandescent 
lamps, one appropriate for illumination of the blueward range of the
detector (at 60 seconds exposure time) and the other appropriate for the 
redward range (5 seconds exposure time).  For wavelength calibration,
spectra of Thorium-Argon lamps were obtained (45 seconds exposure time) 
generally before or after each on-sky group of target exposures, 

Each target observation consisted of between 1 and 3 consecutive exposures.  
The dates for these observations are listed in Tables 1 and 2; multiple
epoch observations are listed separately.  At least 3 stars with precisely
known radial velocities from \citet{nidever02} were observed per observing
run.  

One feature of this instrument is that the optimal focus for the bluest orders 
is not optimal for the reddest orders, so subjective compromises had to be
made in focusing the spectrograph.  Additionally, the spectroscopic focus was 
temperature sensitive, and therefore had to be tuned periodically throughout 
the night to achieve the best spectral resolution.  On the night of 2002 Feb 2, 
many spectra were obtained at an exceptionally poor focus setting.  When poorly
focused, the spectrograph produces spectra that are stigmatic - spectral lines
skewed relative to the dispersion direction which degrades resolution (by an 
amount that is wavelength dependent).  To help assess how poor focus may 
compromise 
the analysis (e.g. $v$sin$i$ measurements), on the night of 2002 Sep 18 we 
observed the radial velocity standard HD 164992 over a broad range of focus 
values.

\section{Spectroscopic Reductions}

\subsection {Image Processing}

Raw data frames were processed using tasks in IRAF and custom IDL 
scripts.  First, the detector
bias was removed.  Although the bias pattern appeared to be stable during
the course of an observing run, the absolute bias level drifted
substantially on time scales of hours, changing by as much as 30\% over the
course of a night.  Bias subtraction was performed, therefore, in a
two-step process.  First, the bias sequence obtained on each night was
median combined using the IRAF ``zerocombine" task, rejecting the highest
and lowest valued bias images.  Second, since the detector bias level
for a given image may have drifted from the median value determined at
the beginning or end of the night, a bias scaling factor was also determined.  
This was done by measuring the median pixel value 
within a specified area 
in the most poorly illuminated, inefficient portion of the array,
located well below 3600 \AA\, where useful spectra 
could not be extracted; even the brightest, bluest stars observed 
on our program showed essentially zero counts above bias in this portion 
of the array.  For each target image, the median bias image from the
beginning or end of the night was scaled by this factor and then
subtracted.

Next, images were trimmed to exclude two unilluminated columns at the 
rightmost edge of the array.  Cosmic ray events were removed using
IRAF's ``cosmicrays" task and fairly liberal parameter settings that
restricted the corrections to only bona fide cosmic rays.

The blue and red flat field sequences from each night were median combined
using IRAF's ``flatcombine" procedure with sigma clipping.  The blue and
red medians were then stitched together approximately 1/3 of the way up
the detector (at line 250) which has the effect of distorting the one
(curved) spectral order which intersects
this line.  IRAF's ``apflatten" task was used to normalize the combined flat
for each night by fitting its intensity along the dispersion using a low
order (typically 3) polynomial while setting all pixels outside an order to
unity.  Flat fielding was performed by dividing each image by the normalized
flat field image using IRAF's ``ccdproc" task.

Cosmetically, the images contain ``tadpole streaks" of high count levels,
comprised of short ``heads" along the dispersion direction
and then decaying count levels along long, mostly single pixel, ``tails" 
perpendicular to the dispersion direction which thus affect multiple orders.  
The number of such defects per image was related to exposure time.
A custom procedure was developed to crawl down each column comparing values 
on either side of the column and thus identifying these features.   The 
pixels in the tail of the streak and a small square around the tadpole head 
were interpolated over.

The close spacing of the spectral orders did not permit measurement or 
removal of scattered light, an effect common in echelle spectrographs.
However, comparisons of our extracted spectra with spectra previously obtained
with other instruments (Section 3.2) and of measured equivalent widths 
with those from the literature (Section 4.1) suggest that the majority of
excess flux caused by scattered light is removed during background 
subtraction.

\subsection {Spectral Extraction and Wavelength Calibration}

One dimensional spectra were extracted from the processed two dimensional
images using IRAF's ``doecslit" task.   Sixty orders were identified over
the 800 x 800 pixel$^2$ array.  A moderate (9th) order trace was employed
and the spectra in each order were summed over 3-5 pixels perpendicular to 
the dispersion.  Variance weighting was used for ``optimal" extraction.
The small separation between orders made removal of background and 
scattered light especially challenging.  After exploring a variety of 
methods for accomplishing this, the most robust method identified was 
simply to subtract the average minimum value within 2 background regions, 
one on each side of the spectrum.  
Aperture definition, background region definition, and order tracing 
were attended as opposed to automated processes.

To test the accuracy of the adopted background/scattered light subtraction,
Jeff Valenti kindly provided a careful comparison of our spectra with spectra 
obtained with the HIRES spectrograph \citep{vogt94}
on the W. M. Keck I Telescope.  The widely spaced orders of HIRES permit 
accurate background and scattered light corrections.  These comparisons
demonstrated that the average scattered and/or background light remaining
in our spectra is approximately 4\% at 3900 \AA, and drops roughly linearly 
to 0\% at 6200 \AA.  The implication is that equivalent widths of features at
blue wavelengths may be diminished by a small amount.  Rather than 
attempt to correct for this in the reduction process, we account for this in 
the relative calibration of properties extracted from blue wavelength features 
(Sections 4.3 and 4.4).

Although multiple Thorium-Argon spectra per night were obtained
(often one for each star observed), in practice we established 
wavelength calibration as follows.  The extraction trace defined by 
a single bright star on each night was applied to its corresponding
Thorium-Argon exposure to extract a one-dimensional wavelength reference 
spectrum.  A wavelength solution was determined using the ``ecidentify''
task in IRAF, and this solution was applied to all spectra for a given
night.  This established wavelength solution was then interpolated onto a 
log-linear scale.

From analysis of spectra taken on the same night,
small spectral shifts of $\sim 0.1$ \AA\, were
common, as inferred from comparisons of the location of telluric absorption 
lines.  These shifts were likely due to flexure of the instrumental optics
with telescope pointing, and consequently resulted in wavelength solution
offsets of this order for many spectra.  As discussed below, this wavelength 
offset is accounted for in determining radial velocities by cross-correlating
telluric features.

\section{Spectroscopic Properties}

\begin{figure*}
\epsscale{1.0}
\plotone{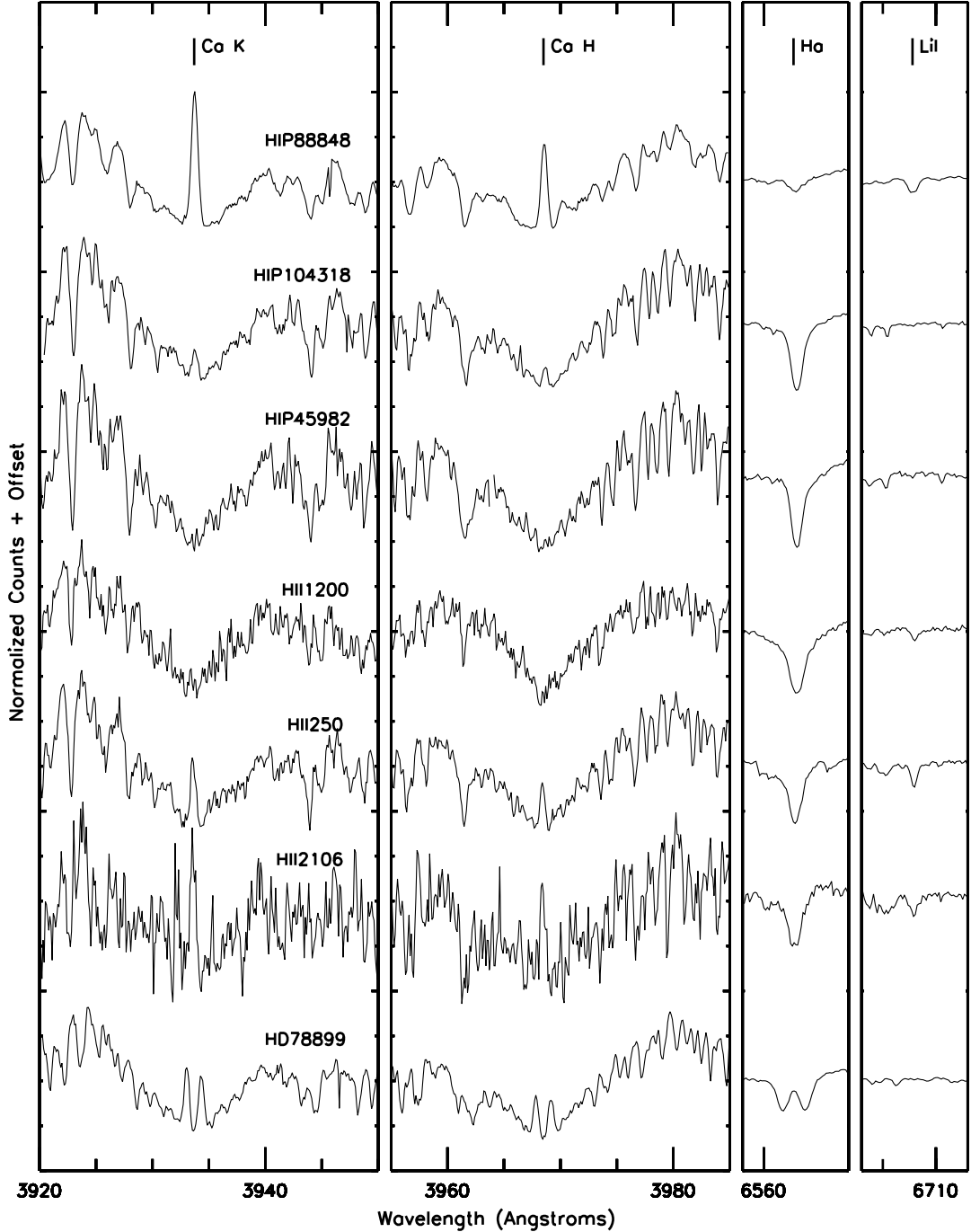}
\caption{Portions of the extracted spectra.  The top 3 spectra are G5 stars with
decreasing levels of Ca\,II H\&K emission (log$R'_{HK} = -4.13, -4.59, -5.01$).  The 
next 3 spectra are Pleiads (and thus coeval), with spectral types of F5 (HII 1200), 
G2 (HII 250), and K0 (HII 2106).  With a SNR of only $\sim 18$ in the Ca\,II region,
HII 2106 is typical of the poorest quality spectra from which log $R'_{HK}$ values can
be extracted; strong core emission can be seen in this case.  The bottom spectrum 
shows the spectroscopic binary HD 78899, whose features are clearly doubled.
\label{fig_spectra} }
\end{figure*}

\begin{figure}
\epsscale{1.0}
\plotone{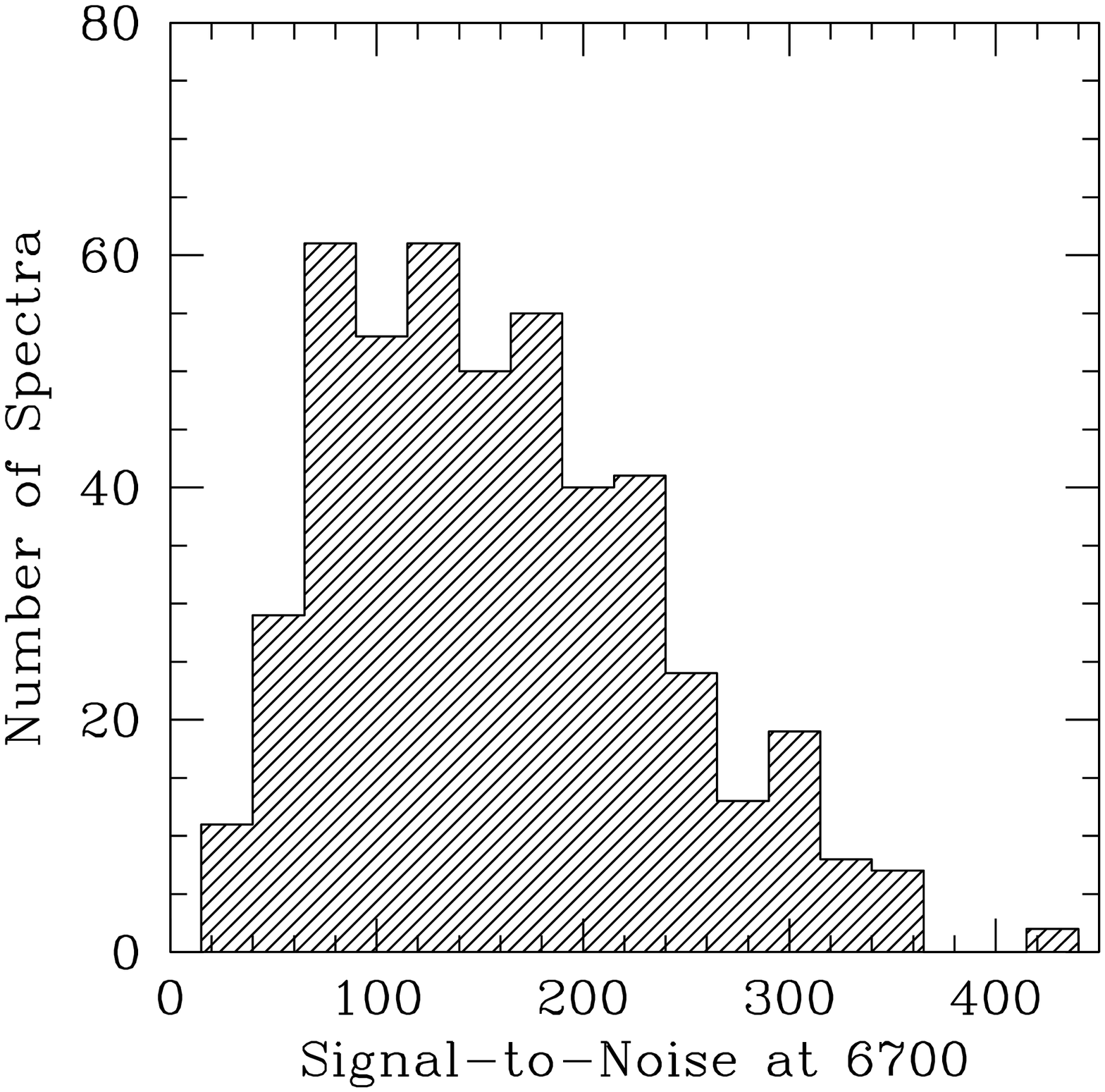}
\caption{Signal-to-Noise distribution of all spectra obtained
(includes multiple measurements and stars observed for calibration) of the 390 
unique targets.
\label{snr_hist} }
\end{figure}

Figure \ref{fig_spectra} shows portions of the echelle spectra for 7 stars near 
the Ca\,II H\&K, Li\,I $\lambda$6708 \AA, and H$\alpha$ features.  
To help assess the quality of the spectra, two signal-to-noise ratios (SNRs) 
are estimated, one at $\sim 6700$ \AA\, and one at $\sim 3950$ \AA\, (labeled
as Li and HK, respectively).  The
SNR at $6700$ \AA\, is determined from the gain-corrected median signal 
within the central portion of this order.  The SNR at $3950$ \AA\, 
is determined similarly, but is based the average of the values determined 
for the two adjacent
orders containing the Ca II H (3968 \AA) and K (3934 \AA) features, and thus
represent an approximate average over the continuum and broad absorption.
These estimates are valid above values of $\gtrsim 20$, in which the noise
is dominated by photon statistics.  Lower values may actually over-estimate 
the SNR, because of the increased relative error in the background, scattered
light and bias level subtractions (Section 3.1).  The SNR distribution at 6700 \AA\, 
is shown in Figure \ref{snr_hist}.

The following sections described the procedures used to extract stellar properties
from these spectra.  In comparisons
with previous measurements, values determined in this study are referred to as P60
values.  Specifically, equivalent widths (EWs) of Li\,I $\lambda$6708 \AA\, 
and H$\alpha$ are measured in Section 4.1, radial and rotational velocities 
are determined in Section 4.2, temperature and surface gravity indices are
discussed in Section 4.3, and measured S values and calculated $R_{HK}^{'}$
indices are presented in Section 4.4.

\subsection{Equivalent Widths of Li\,I 6708 \AA\, and H$\alpha$}

Equivalent widths of the Li\,I $\lambda$6708 \AA\ doublet ($\lambda$6707.76, $\lambda$6707.91), 
which is spectroscopically unresolved in our observations, 
and H$\alpha$ ($\lambda6562.8$ \AA) are measured for all spectra using the $splot$ 
task within $IRAF$.  For measurement of the Li\,I absorption feature which is located 
near the peak of the blaze function in these spectra, the local continuum is determined 
by fitting a legendre polonomial to a $\sim 30$ \AA\, portion of the spectrum, excluding 
values less than 2$\sigma$ below the continuum (e.g. stellar absorption features)
or more than 4$\sigma$ above the continuum (e.g. noise spikes) in this
fit.  The order of the polynomial is varied (typically from 4 to 7) to best
match the local continuum.  With the continuum
defined in this way, the EW values are determined by fitting a Gaussian 
profile to the absorption feature.  Multiple measurements on a single
night are averaged, while separate epochs are listed independently.
If no feature is discernable in the spectrum, upper limits 
are determined from the noise in the local continuum.  Table 1 lists 
the measured values.

The measured Li\,I EW values include a contribution from an Fe\,I line at 
6707.441 \AA.  \citet{soderblom93} find that the strength of this temperature
sensitive Fe\,I 
feature varies with $B-V$ color as EW[Fe\,I $\lambda$6707.441] = 20($B-V$) - 3 m\AA.
Since the $B-V$ values of most stars are known or can be estimated (Section
4.4), the contribution of this Fe\,I line to the Li\,I EW values can be
removed.

\begin{figure}
\epsscale{1.0}
\plotone{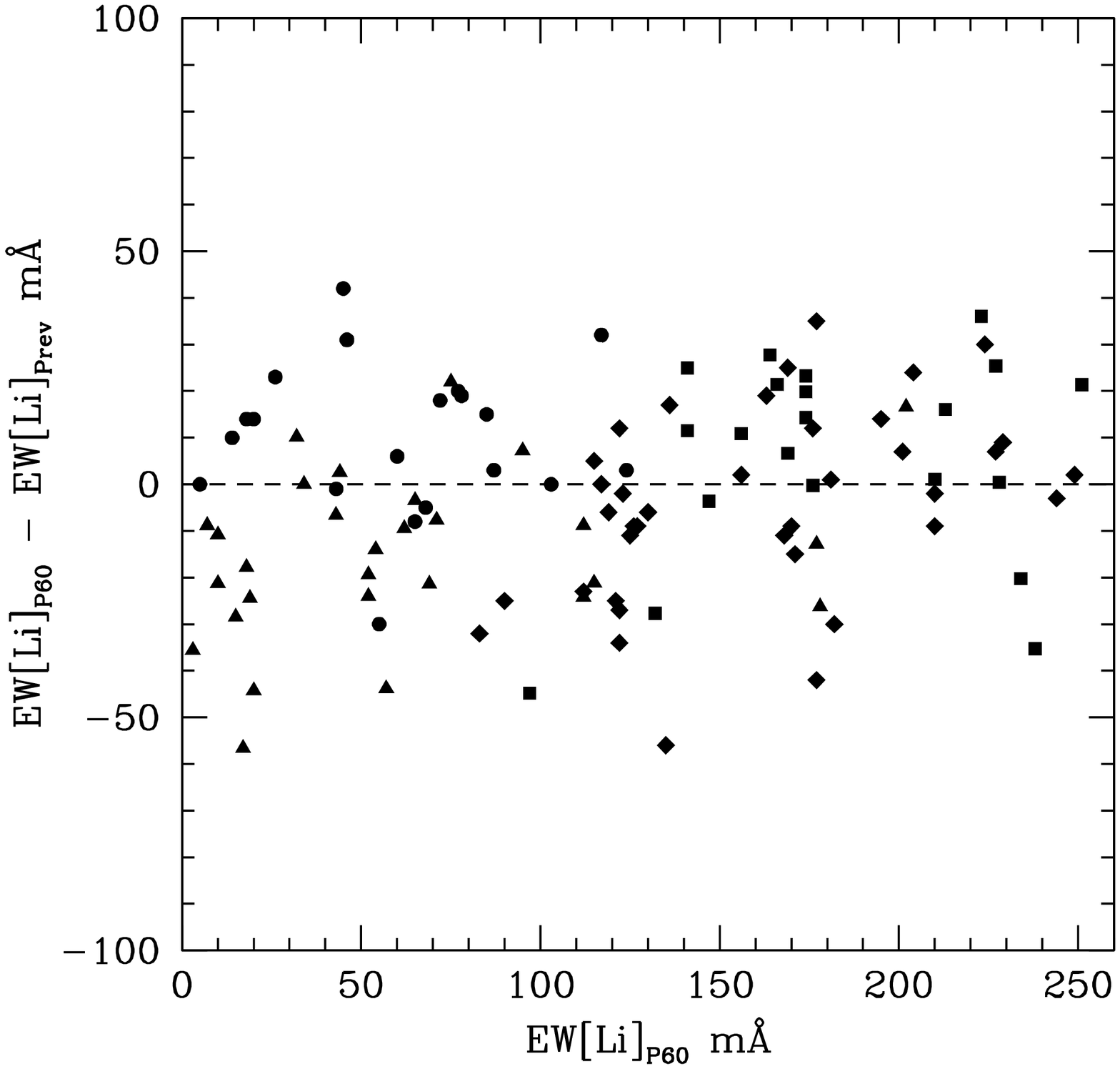}
\caption{Comparison of measured EW[Li\,I $\lambda$6708 \AA] values with values from
\citep[][$squares$]{soderblom93},
\citep[][$circles$]{thorburn93},
\citep[][$diamonds$]{wichmann03}, and
\citep[][$triangles$]{strassmeier00}.
\label{lithium} }
\end{figure}

To assess possible systematic uncertainties caused by the subjective aspects
involved in measuring EW values (e.g. continuum determination) or instrumental
effects, we compare our measurements to previous
values determined from high dispersion spectroscopy.  Of the
spectroscopically single stars observed here, 21 Pleiads were observed by  
\citet[][$R=50,000$]{soderblom93}, 19 Hyads were observed by
\citet[][$R=26,000 - 32,000$]{thorburn93}, 26 field stars were
observed by \citet[][$R=40,000$]{wichmann03}, and 28 field stars were
observed by \citet[][$R=25,000$]{strassmeier00}.  The uncertainties in the 
EW[Li]s from these previous studies are typically 5-10 m\AA.  Of these 94 
stars, 19
were observed twice by us, yielding a total of 113 measurements for
comparison.  Figure \ref{lithium} shows the difference between the P60 and
previous EW values; all values are as measured and thus uncorrected for
the contamination of the Fe\,I line.  Overall the agreement is good.  The
average and/or median difference for any of the 4 studies is $\pm 15$ m\AA,
with standard deviation of differences that range from 16 - 21 m\AA.  The
average difference with the entire comparison sample is -8 m\AA\, with a
standard deviation of 20 m\AA.  We conclude that any systematic bias in our
measurements is $\lesssim 10$ m\AA.  We adopt a uniform uncertainty 
of 0.02 \AA.

For H$\alpha$ EWs, considerable care had to be taken both in normalizing to
the local continuum and in fitting a functional form to the line profiles 
because of the broad wings. The local continuum was determined by fitting 
a legendre polonomial generally of order 10 to the entire spectral order, 
excluding values less than 2$\sigma$ below or more than 4$\sigma$ above 
the continuum.  The procedure varied slightly as dictated by the noise
level in the data, by the line breadth (because of, for example, rapid rotation),
or if H$\alpha$ was found to be in emission.  Both the sigma-rejection 
limits and the fitting order were modified in these cases to optimize
the continuum fit.  The EW values were determined by fitting a Voigt
profile to the absorption feature; in a few cases of dramatically filled
in H$\alpha$, direct integration was used to determine the EW
rather than function fitting.  EWs are listed in Table 1; negative values indicate
H$\alpha$ emission.

\begin{figure}
\epsscale{1.0}
\plotone{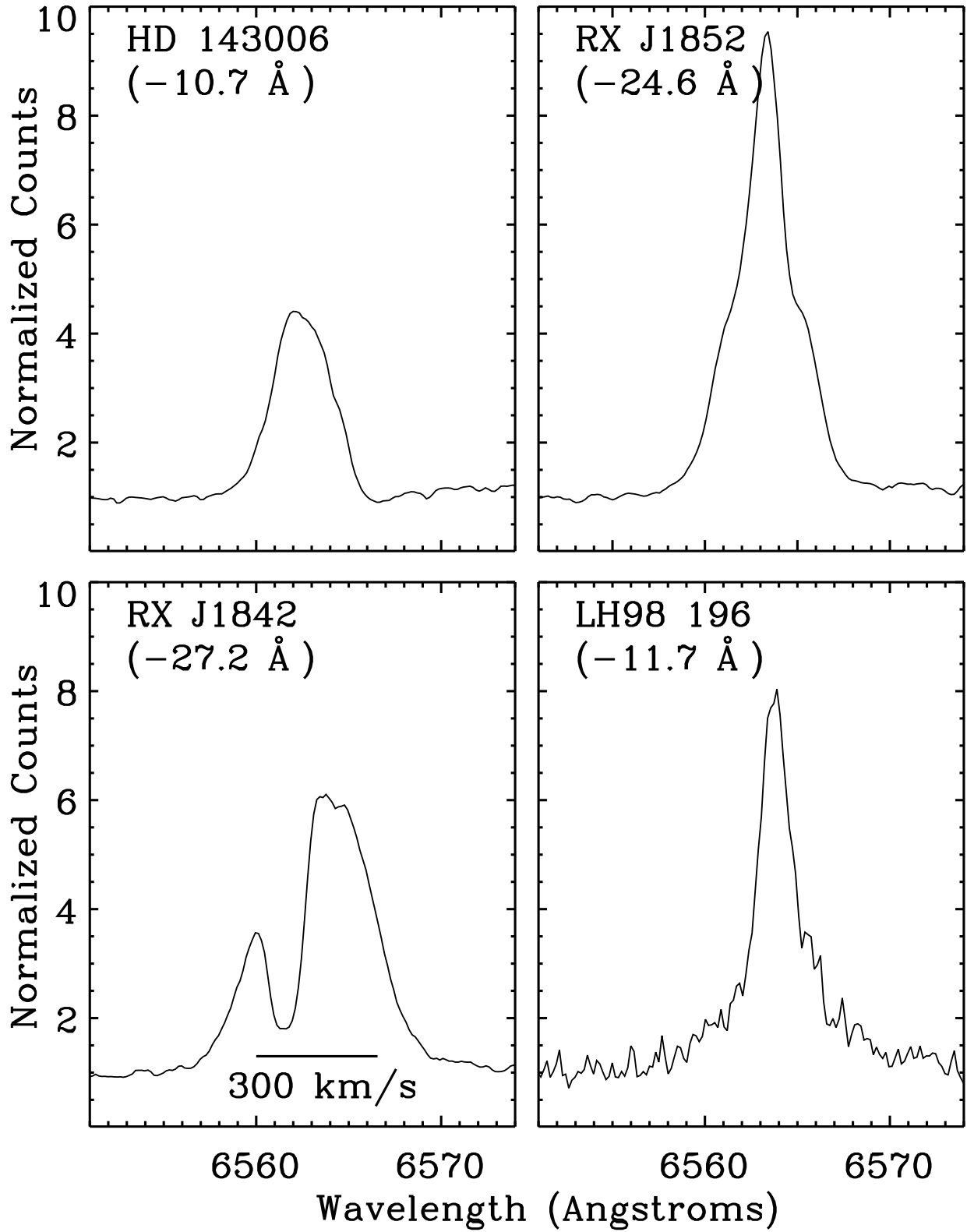}
\caption{H$\alpha$ profiles for the 4 strong emission-line stars normalized by 
the local continuum; the H$\alpha$ equivalent width for each star is given in 
parentheses.  A velocity width of 300 km s$^{-1}$ is indicated.
\label{fig_halpha} }
\end{figure}

Four stars (HD 143006, RX J1842.9-3532, RX J1852.3-3700, LH98 196) show 
especially strong H$\alpha$ emission ($> 10$ \AA); their H$\alpha$ profiles are 
shown in Figure \ref{fig_halpha}.  In these cases, the strength and breadth of 
these emission-line profiles are greater than that expected from chromospheric 
activity alone, and more consistent with that expected from the accretion of 
circumstellar material \citep[e.g.][]{muzerolle98, wb03}.  We suggest that these 
stars are accreting.

For the 8 stars identified as double-lined spectroscopic binaries, EW[Li\,I]s 
and EW[H$\alpha$]s for the individual components were determined by fitting 
two Gaussian profiles to the doubled features, when possible.  For these 
pairs, we also measured EWs of the temperature sensitive Ca\,I 
$\lambda$6717 \AA\, 
feature, which can be used to identify better the primary and the secondary.  
The EW values for the binaries are listed in Table 2.

\subsection{Radial and Rotational Velocities}

Radial velocities and projected rotational velocities are determined via a
cross-correlation analysis similar to that of \citet{wh04}.  The analysis is
restricted to the 36 orders spanning from $\sim$ 4000-7500 \AA, excluding 
the 3 orders containing the telluric B-band, H$\alpha$, and Na\,II D features; 
orders at longer wavelengths have poorer wavelength solutions and those at
shorter wavelengths
typically have low SNRs.  From this restricted range, orders having an 
average SNR greater than $\sim 25$ are then cross-correlated with the spectral
orders of at 
least 3 slowly rotating comparison standards that have radial velocities 
accurate to 0.3 - 0.4  km s$^{-1}$ \citep{nidever02}.  The radial velocity, relative 
to each standard, is then determined from the average velocity offset 
measured from all orders.

To determine the heliocentric radial velocities of the stars, the relative
radial velocity must be corrected for barycentric motions and possible
errors in the wavelength solution of either the standard or 
target spectrum.  The former is determined using the \textit{rvcorrect}
task in IRAF.  The latter correction is determined by cross-correlating 
the telluric A- and B-bands of the standard and the target star; the 
telluric correlations would yield an offset of zero for perfect
wavelength calibration, but are typically 1-3 km s$^{-1}$ for the spectra
analyzed here.  The multiple heliocentric radial velocity estimates,
determined from multiple radial velocity standards, are then averaged
to find the final radial velocity.

\begin{figure}
\epsscale{1.0}
\plotone{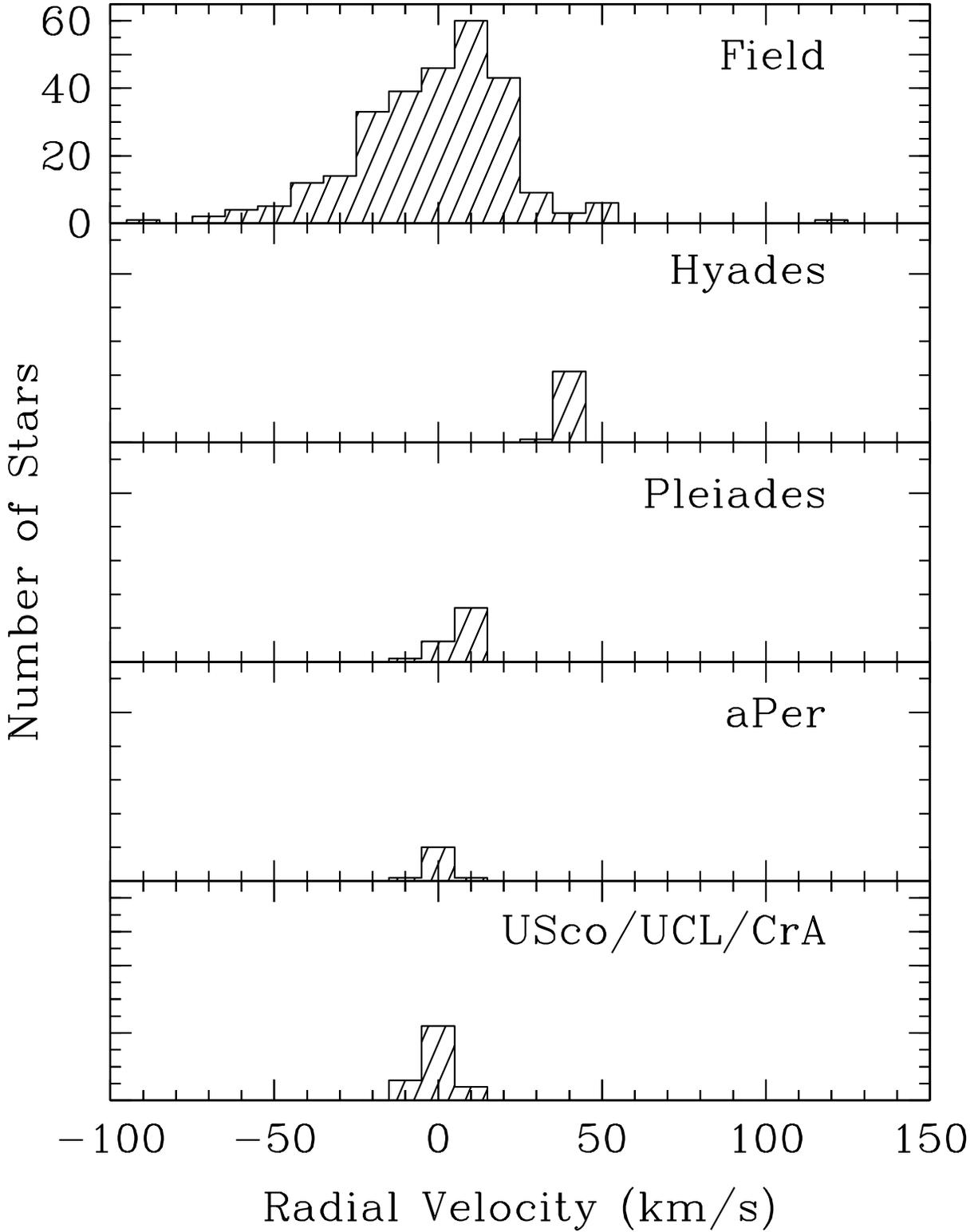}
\caption{Distributions of radial velocities of the observed sample.  The top 
panel shows the distribution for field stars and the bottom four panels show
the distributions for cluster stars.
  \label{fig_rv_hist} }
\end{figure}

Uncertainties in the radial velocities are estimated from a combination of
statistical and empirical error estimates.  First, the uncertainty in
the radial velocity relative to each stardard is assumed to be the
uncertainty in the mean value of all orders used, combined with the 
uncertainty in the radial velocity of the standard (typically 0.3 km 
s$^{-1}$).  The radial velocity estimates from all standards are then
combined using a weighted average; the resulting statistical 
uncertainties are typically small (0.1 - 0.3 km s$^{-1}$).  As an
empirical check on these uncertainty estimates, the radial velocity of each
standard is determined from all other radial velocity standards observed
during that run and
compared to its assumed value.  The average agreement of the radial
velocity standards is typically a couple times larger than the statistical
uncertainties, except in low SNR cases where the statistical uncertainty
dominates.  This suggests there may be systematic effects unaccounted for
in the statistical uncertainty estimates.  Thus, we combine the statistical
uncertainty with this empirical uncertainty to determine final
(conservative) radial velocity uncertainty estimates given in Table 1.  
Figure \ref{fig_rv_hist} shows the distributions of radial velocities for the
field, open cluster, and young cluster samples.  

Rotational velocities ($v$sin$i$) for the observed targets are determined
from the width of the peak in their cross-correlation with a slowly
rotating standard.  Specifically, the peak is fit by a parabola with the
fitted width at half-peak maximum.  To
convert this width to a $v$sin$i$ value, an empirical width versus
$v$sin$i$ relation is determined by cross-correlating many artificially ``spun-up''
spectra with the same slowly rotating standard.  These artificially broadened spectra
are constructed using the rotational profiles given in \citet[][$\epsilon = 0.6$]{gray92}.
This procedure is computed for each order, yielding a $v$sin$i$ estimate for
each order.  For each standard used, a
$v$sin$i$ value is computed from the median of all order estimates; a median is
preferred since the $v$sin$i$ measurements are often upper limits.

Uncertainties in $v$sin$i$ are determined first from the standard deviation
of the multiple $v$sin$i$ measurements provided by all standards used.  For
observations consisting of multiple exposures, the uncertainty is the
average standard deviation of these measurements.
The $v$sin$i$ values are assumed to be upper limits if less than 10 km s$^{-1}$
(just over one-half the velocity resolution), or if $v$sin$i$ 
is less than 3 times the uncertainty in $v$sin$i$.
We find that all templates of F, G, and K spectral types give consistent
$v$sin$i$ values; an exact match in spectral type is not critical in
determining $v$sin$i$.

As noted in the introduction, poor focus plagued some of our measurements,
especially on the night of 2002 Feb 2.  To assess how this may affect
inferred properties, the above radial velocity and rotational velocity
analysis was conducted on the observations of radial velocity standard HD
164922 observed at multiple focus settings (Section 2).  Fortunately the
inferred radial velocities for modest focus offsets, typical of changes
over a given night, agree with the nominal value to $\sim 1$ km s$^{-1}$,
consistent with the inferred uncertainties.  For extreme focus offsets,
however, the discrepancy can be as large as 3 km s$^{-1}$.  Similarly, for modest
focus offsets from nominal, the inferred $v$sin$i$ values would increase modestly, but
since the standard deviation of measurements among the orders 
correspondingly increased,
the end result is simply larger upper limits (e.g. $<$ 16 km s$^{-1}$ for HD
164922, compared with $< 10$ when observed at nominal focus).  
However, for extreme focus offsets, the spectrsopic blurring
resulted in a large apparent $v$sin$i$ (e.g. $22.0 \pm 3.3$ km s$^{-1}$ for HD
164922).  Since the majority of our spectra have at most only modest focus
errors, no correction for this is made.  Rotational
velocities inferred from observations made on 2002 Feb 2, however, are 
marked with a colon in Table 1 as likely being artificially too large.

\begin{figure}
\epsscale{1.0}
\plotone{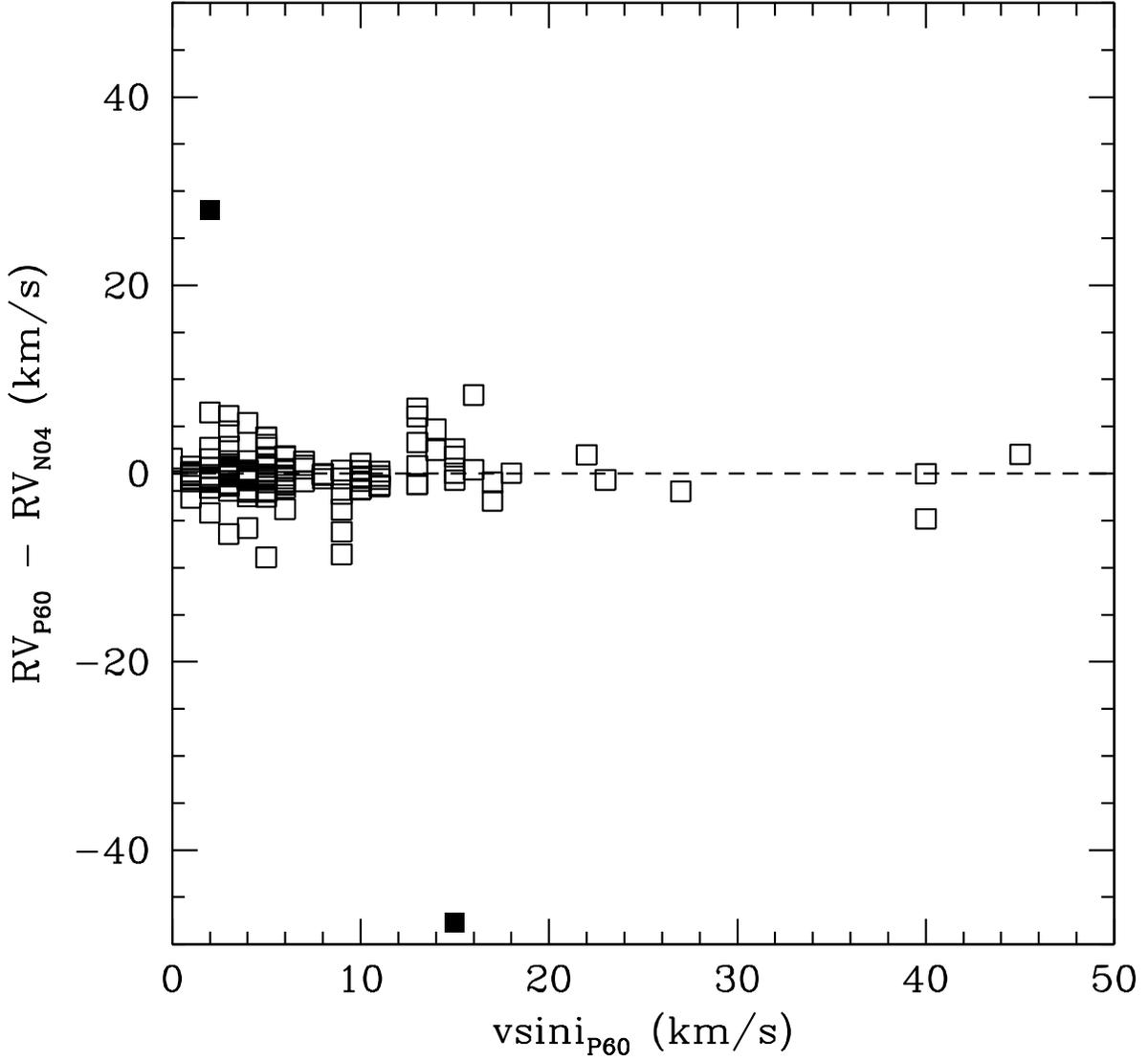}
\caption{Difference between P60 radial velocities and values
  reported by \citet{nordstrom04}.  The \textit{solid squares} are
  likely spectroscopic binaries.  \label{fig_rv_compare} }
\end{figure}

To empirically assess systematic errors in our radial and rotational
velocity measurements, our values are compared with values measured by 
\citet[][which appeared after all of our data were taken]{nordstrom04}.  
There are 135 stars in common between the surveys
that (1) are not previously identified spectroscopic binaries, (2) are not used
as radial velocity standards by us, and (3) have radial velocity
uncertainties less than 4 km s$^{-1}$.    
Figure \ref{fig_rv_compare} shows the
difference between our radial velocities and those of \citet{nordstrom04}.
The agreement is quite good; the average difference is -0.2 km s$^{-1}$
with a standard deviation of 2.4 km s$^{-1}$.  With the exception of 2 stars,
HIP 67904 and HIP 61072, all measured radial velocity values 
agree to within 10 km s$^{-1}$. HIP 67904 and HIP 61072, on the
other hand, have significantly different radial velocity measurements ($>
25 \sigma$ difference).  As discussed below, we suggest these stars are
spectroscopic binaries.  This comparison also demonstrates that the radial 
velocities of stars with $v$sin$i$ values as large as 45 km s$^{-1}$
can be measured without significantly degraded precision.

\begin{figure}
\epsscale{1.0}
\plotone{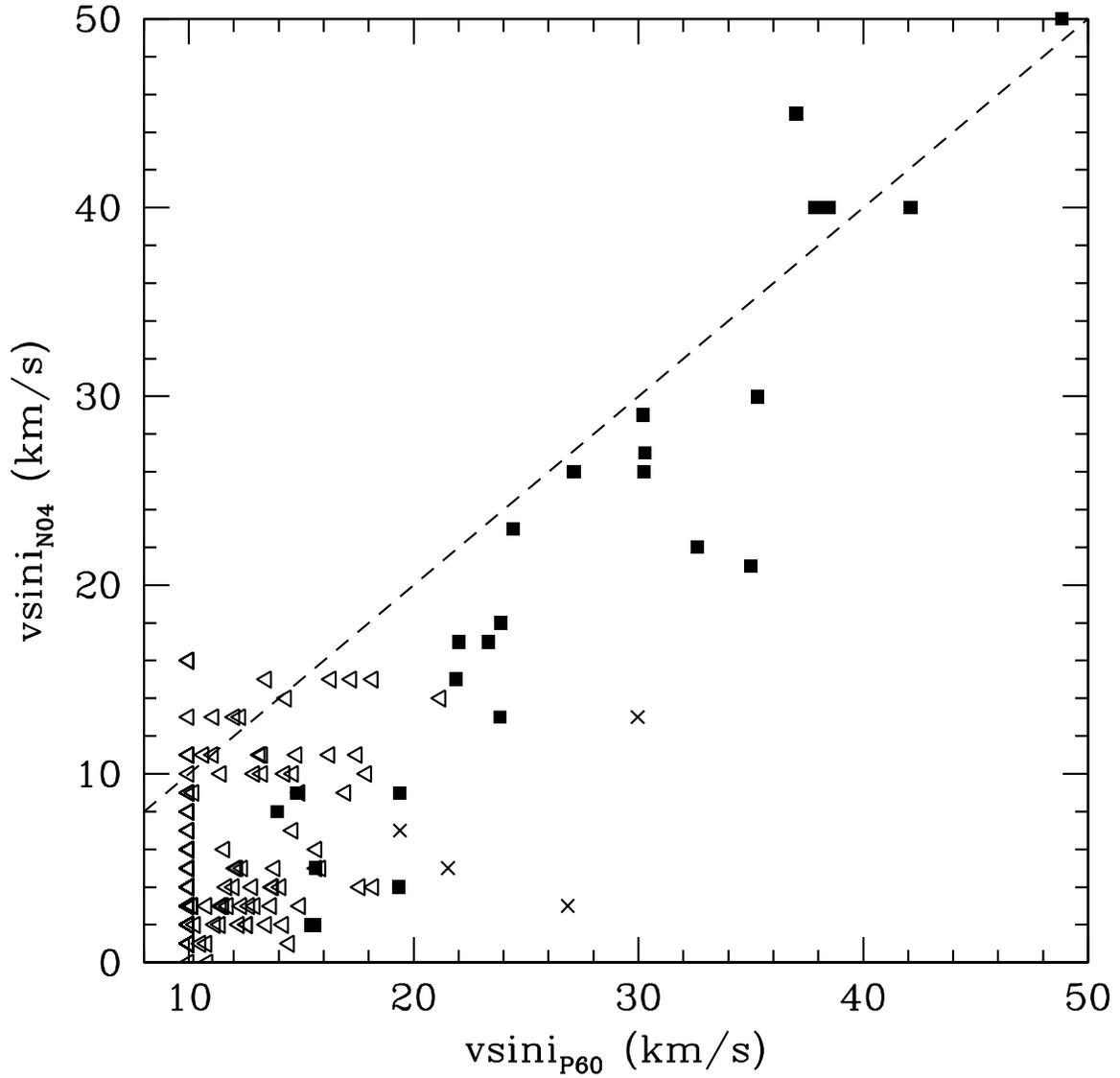}
\caption{Comparison of measured $v$sin$i$ values with values from 
  \citet{nordstrom04}.  \textit{Triangles} are upper limits and
  \textit{squares} are measured values; \citet{nordstrom04} report no upper
  limits.  The 3 x's indicate measurements from the night of 2002 Feb 2,
  which had significant focus problems (Section 4.2).
\label{fig_vsini_compare} }
\end{figure}

In Figure \ref{fig_vsini_compare} are shown \citet{nordstrom04} $v$sin$i$
values versus the P60 $v$sin$i$ values for the overlapping sub-sample
described above.  While \citet{nordstrom04} report $v$sin$i$ values
as low as 0 km s$^{-1}$ (which is likely unrealistic), our lower resolution 
data restrict us to $v$sin$i$ upper limits of 10 km s$^{-1}$ or greater.  The 
majority of our $v$sin$i$ upper limits are thus greater than the values measured 
by \citet{nordstrom04}, which inhibits interpretation of an overall comparison.  
However, a comparison of the 22 stars with measured 
values, however, suggests a systematic difference.  The P60 $v$sin$i$ values are
larger, in the median, than those of \citet{nordstrom04} by 8.6 km s$^{-1}$.  
A portion of this difference 
may be a consequence of the spectroscopic blurring caused by the focus errors 
discussed above.  When the 3 measured values from 2002
Feb 2 are removed, the P60 values are then larger by only 6.3 km s$^{-1}$, in the
median.  However, an independent comparison of the 389 stars with $v$sin$i$ 
measurements in both \citet{nordstrom04} and \citet{strassmeier00}, upper
limits excluded, indicate the \citet{nordstrom04} values are less than the
\citet{strassmeier00} values by 2.6 km s$^{-1}$.  Thus, a portion of the  
difference with our measurements may be a consequence of artificially low values 
reported by \citet{nordstrom04}.  Overall we conclude that the majority of our 
measured $v$sin$i$ values suffer from no systematic biases greater than $\sim
5$ km s$^{-1}$.  The exceptions are stars observed on 2002 Feb 2, which may 
be biased toward artificially large values by $\sim 10 - 20 $ km s$^{-1}$ because
of especially poor focus.  Since we are uncertain what the systematic 
correction to the $v$sin$i$ values for these stars should be, if any, we make 
no correction for this in the values presented in Table 1.

\subsection{Temperature and Gravity Indicators}

\begin{figure}
\epsscale{1.0}
\plotone{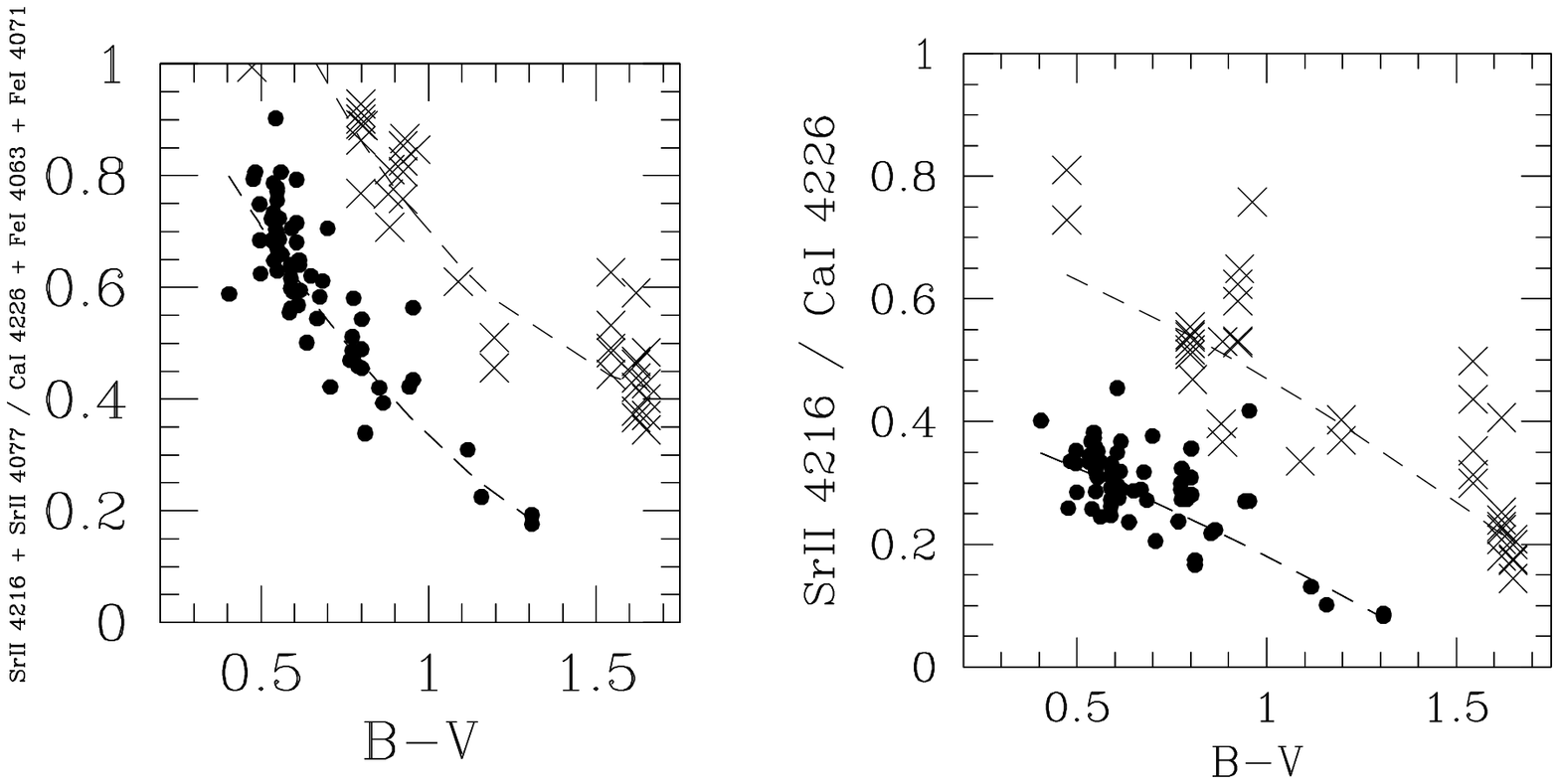}
\caption{Gravity sensitive equivalent width ratios 
versus B-V color.  Dwarf stars are indicated with \textit{solid 
circles} and giant stars are indicated with \textit{crosses}.  \label{fig_gravity} }
\end{figure}

Within the broad wavelength coverage provided by our echelle spectra are 
many temperature- and gravity-sensitive lines.  Empirically, we find that the
EWs of the photospheric features Fe\,I $\lambda$4063, Fe\,I $\lambda$4071,
and Ca\,II $\lambda$4226 \citep[see][]{jaschek87} are especially useful
temperature diagnostics.  In our data, these features all
increase in strength in a roughly linear fashion from spectral types F8 to K5, 
before degenerate flattening and turnover occurs towards later spectral types.
However, fits of the EWs versus $B-V$ color for main sequence stars (not 
shown here)
are accurate to only 0.05-0.1 in $B-V$.  This implies that if we were to attempt
spectral classification from the EWs, spectral type uncertainties of more than one
spectral subclass would result.  Therefore
we do $not$ use these features to estimate spectral types.  Spectral types 
instead are taken from several catalogues of stellar spectral classification, 
when available.  These, in order of
preference are Houk \& Swift (1999), Skiff (2005), Kharchenko (2001),
Kharchenko, Piskunov, Scholz (2004).  The spectral types are listed in
Table 1; 349 of the 370 stars observed have stellar spectral types.

For gravity indicators, we considered the ratios of the gravity sensitive Sr\,II 
$\lambda$4216 and $\lambda$4077 lines \citep[see][]{jaschek87} to the above 
temperature sensitive lines.  In Figure \ref{fig_gravity} we show for our dwarf 
and giant standard stars the ratios (Sr\,II $\lambda$4216 + $\lambda$4077)
to (Ca\,I $\lambda$4226 $+$ Fe\,I $\lambda4063$ + Fe\,I $\lambda4071$) and Sr\,II 
$\lambda$4216 to Ca\,I $\lambda$4226, as a function of $B-V$ color.  These
ratios are sensitive to both temperature and gravity; the dwarf
and giant stars are reasonably well separated in these plots.  Parabolic fits 
to the dwarf and giant points are shown.  The majority  of our target 
objects cluster around the dwarf locus, though some are securely 
between the dwarf and giant loci.  Other diagnostic figures, such as
Sr\,II $\lambda4077$ / (Fe\,I  $\lambda4063$ +  Fe\,I $\lambda4071$) 
versus Ca\,I  $\lambda4226,$ (which has the advantage of considering only 
our spectroscopic data), have similar appearances.  In principle one might 
use such figures to quantitatively estimate the surface gravities of our program
stars; in practice, however, the data do not allow this to be
carried out with the precision normally expected of surface gravity
measurements.  Rather than estimate surface gravities, we simply identify
all stars which are more than the 1$\sigma$ above the dwarf loci in both 
diagnostic planes.  Sixty-six potentially low surface gravity stars are 
identified by these criteria and marked in Table 1.

\subsection{Chromospheric Ca II H and K Activity Measurements}

Core emission in the Calcium H and K spectral lines serves as an indicator
of stellar chromospheric activity, which is known to be correlated with 
stellar age \citep[e.g.][]{soderblom91}, primarily as a consequence of rotation.  
We parameterize the core emission
in our spectra using the standard Mount Wilson Project procedures
\citep{wilson68, vaughan78}.  Our prescription for
measuring the core emission closely follows that outlined by
\citet{duncan91}.  First we define the index of core emission, $S,$ as
\begin{equation}
\label{eq:S}
S = \frac{N_H + N_K}{N_R + N_V},
\end{equation}
where $N_i$ is the number of counts in band $i$.  The continuum bands, $R$
and $V,$ are $20$ \AA\,  wide and centered at $4001.07$ \AA\,  and $3901.07$ \AA,
respectively.  The emission bands, $H$ and $K,$ are centered on the
emission features whose natural centers are at $3968.470$ \AA\,  and
$3933.664$ \AA, respectively.  The bandpass for the $H$ and $K$ channels is
triangular, with a full width at half maximum of $1.09$ \AA.
Before measuring the counts within these four bands, the spectra are
normalized to account for the blaze function and shifted to correct
for radial velocity.  The spectra are normalized in a two step iterative
process.  First, a $\sim 4$th order polynomial is fit to 
the spectra; the broad $H$ and $K$ absorption features are excluded in
this fit.  A second polynomial
is then fit to the spectra, excluding all spectral features that extend 
0.5 $\sigma$ below the initial polynomial fit.  The spectra are then normalized
by dividing by this polynomial.  The velocity shift is determined 
by cross-correlating the 2 spectra orders with the Ca II H and K
features, with the spectrum of a bright comparison standard.  The implied
radial velocity is compared with that determined in Section 4.2 to ensure 
consistency; for the chromospheric activity analysis we adopted values
determined directly from these orders as they typically 
provide the most accurate shift for these orders.

With velocity shifted and normalized spectra, the counts within 
the continuum and emission bands of the S index are summed.  
For the emission bands, 
the number of counts in each pixel is first multiplied by a number from 
0 to 1 to mimic a triangular bandpass with a full-width at half-maximum 
of $1.09$ \AA.  With these sums, equation \eqref{eq:S} is used to compute 
$S$.  


To correct for systematic offsets in the measured $S$ values relative to 
previous work, possibly due to slight differences in the methodology or data
quality, the $S$ values are compared to those 
measured by \citet[][also appearing after all data
for this program were gathered]{wright04}.  These comparisons 
permit transformation of the $S$ values to what we
refer to as Mount Wilson $S$ values, $S_{MW}$.  This is realized as a first 
order regression analysis on log $S$ in the form of
\begin{eqnarray}
\log S_{MW}        & = & A + B\log S.
\label{eq:Smw}
\end{eqnarray}
The uncertainty, $\sigma_{S_{MW}}$, is given by the
standard formula 
\begin{equation}
\label{eq:sigmasquared}
\sigma_{S_{MW}}^2 = \left( \frac{\partial S_{MW}}{\partial A} \sigma_A
\right)^2 + \left( \frac{\partial S_{MW}}{\partial B} \sigma_B \right)^2.
\end{equation}

The analysis was performed on 8 of the 9 
observing runs and for the entire overlapping sample; there were
too few overlapping stars for the 2001 December observing run for comparison.
Fortunately, the uncertainties in the implied correction estimates for 
7 of the 8 observing runs agree well with that determined for the entire 
overlapping sample.  The one exception was for the July 2001 observing
run, for which the corrections appeared significantly different.  Therefore,
the corrections determined for this run were used to transform its $S$ 
values to $S_{MW}$ values, while the remaining runs used the corrections
determined for the entire sample, which are more precise.  The resulting Mount
Wilson Calcium index values and uncertainties are listed in Table 1.  
Stars with uncertainties that are possibly much larger than reported, because
of a large systematic uncertainty that is not formally propagated (e.g. radial 
velocity error) are marked with an "e" in Table 1.

Finally, the $S_{MW}$ values are converted to $R'_{HK}$ values, which express
the activity as a fractional ratio with the underlying star, following
the prescription described in \citet{noyes84}.  However, this conversion
requires proper characterization of the underlying photospheric flux, which
can be assessed from the intrinsic stellar $B-V$ color.  $B-V$ colors for most
stars are obtained from the Tycho-2 catalogue \citep{hog00}.  These values are
transformed to Johnson $B-V$ colors, following the relations in
\citet{mamajek02}\footnote{There is a typographical error in the relations 
of \citet{mamajek02}.  The +0.007813 in equation C6 should be -0.007813.}.
For objects without Tycho photometry, the $B-V$ colors are estimated from the
stellar spectral type (Section 4.3) following the relations of \citet{johnson66}.

$R'_{HK}$ is calculated as the difference in the logarithms of the flux
calibrated core emission, $R_{HK}$, and photospheric emission, $R_{phot}$:
\begin{equation}
\label{eq:R'_HK}
R^{\prime}_{HK} = R_{HK} - R_{phot}.
\end{equation}
Here $R_{HK}$ is calculated from the $S_{MW}$ values and calibrated using
the stellar $B-V$ color, and $R_{phot}$ is calculated from the $B-V$ color
alone (see Appendix b in Noyes et al. 1984 for explicit formulae).  Although the 
calibration of these terms is determined over the restricted $B-V$ range of 
$0.44<(B-V)<0.82,$ we use it for the full $B-V$ range of the observed sample 
($0.19 <(B-V)<1.65$).  The resulting $R'_{HK}$ values 
are listed in Table 1.

\section{Sample Properties}

\subsection{Spectroscopic Binaries}

Table 2 lists the 8 stars with photospheric spectral features that are doubled 
in our observations.  Two of these double-lined spectroscopic binaries were 
previously known (HE 848, SAO 178 272), while HD 78899, TYC 7310 503 1, 
TYC 7305 380 1, and HD 140374 are newly identified.  The multiple epoch 
observations of some stars also allows us to identify single-lined binaries based 
on changes in 
radial velocity.  We consider an object a candidate binary if its radial velocity 
changed by more than 10$\sigma$ between observations; 3 stars meet this criteria:
HD92855 (HIP52498), HD132173 (HIP73269), HD61994 (HIP38018; see Table 1).  
Similarly, based on comparisons with radial velocities published in \citet{nordstrom04}, 
HD108944 (HIP61072) and HD121320 (HIP67904) are candidate binaries (see Figure
\ref{fig_rv_compare}).  All 5 are newly identified candidate binary systems.  Finally, since the 
radial velocity dispersion in young clusters is typically $< 2$ km s$^{-1}$, comparison of 
cluster member radial velocity measurements
with the median cluster values can identify
single-lined spectroscopic binaries.  Based on these comparisons, the two Pleiades 
members HII 571 and LH98 103 are identified as spectroscopic binaries; both 
have been identified previously as such \citep{mermilliod92, queloz98}.

\subsection{Sample Age Dispersion}

Several of the quantities extracted from the observed spectra are  likely
to be correlated with stellar age, based on extensive studies in the literature. 
Generally speaking, younger stars tend to be more rapidly rotating, have 
larger lithium abundances, and have stronger chromospheric emission lines 
such as Ca\,II H\&K and H$\alpha$.  Thus the measured quantities 
$v$sin$i$, EW[Li\,I $\lambda$6708], log $R'_{HK}$, and EW[H$\alpha$] 
should correlate with age.  Unfortunately the precision with which stellar ages 
can be estimated from these quantities is in general poor, in part because of
the additional dependencies upon stellar spectral type and rotation.
Rather than attempting to disentangle these complicating effects for individual 
stars, we simply present the data to illustrate the empirical correlations 
amongst the measured quantities and to assess the overall age range of the 
observed sample.

\begin{figure}
\epsscale{1.0}
\plotone{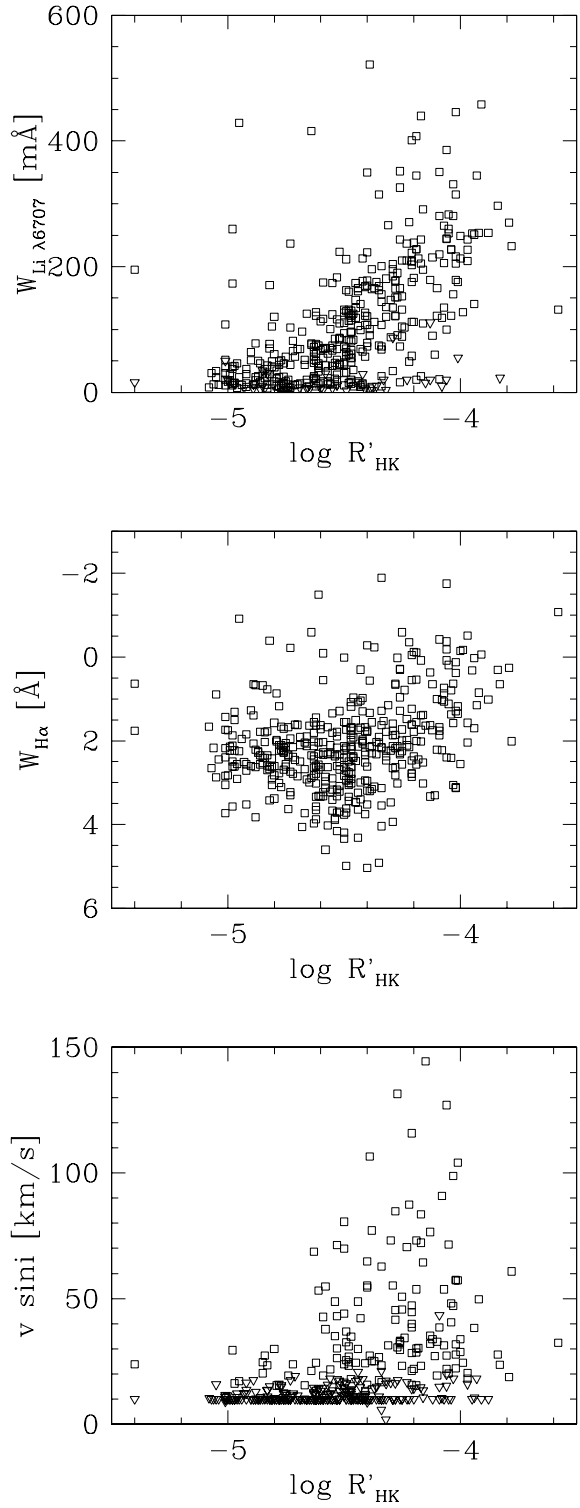}
\caption{EW[Li\,I 6708], EW[H$\alpha$] and $v$sin$i$ versus the chromospheric 
activity diagnostic log\,$R'_{HK}$.  Measured values are shown as \textit{squares} 
while upper limits are shown as \textit{triangles}.
\label{fig_rphk_age} }
\end{figure}

\begin{figure}
\epsscale{1.0}
\plotone{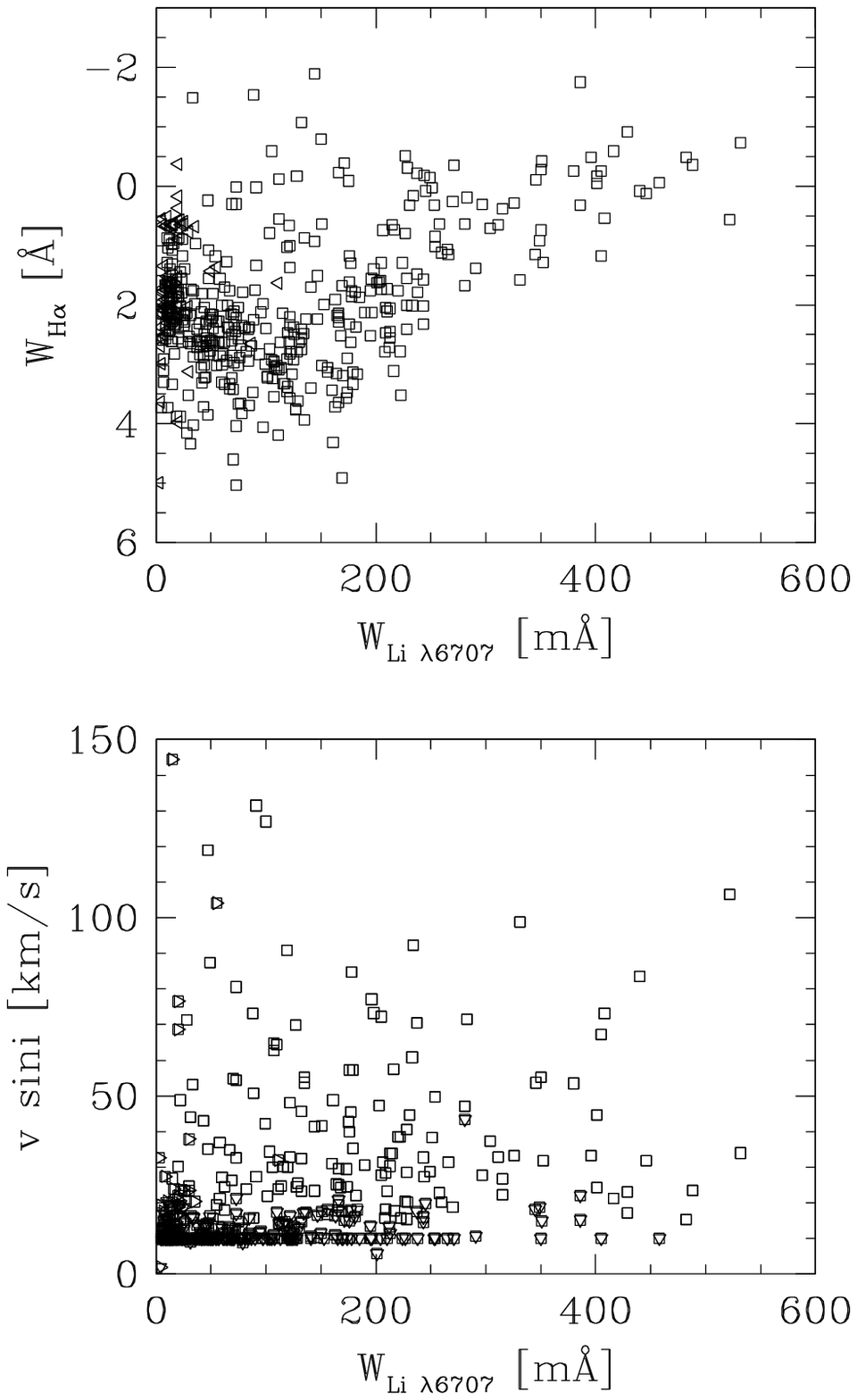}
\caption{EW[H$\alpha$] and $v$sin$i$ versus the EW[Li\,I 6708].  Measured values 
are shown as \textit{squares} while upper limits are shown as \textit{triangles}.
\label{fig_lithium_age} }
\end{figure}

Figure \ref{fig_rphk_age} shows the distributions of EW[Li\,I 6708], EW[H$\alpha$] 
and $v$sin$i$ versus log\,$R'_{HK}$.  Error bars are 
not displayed due to crowding, but can be inferred from the information in Table 1.
The EW[Li\,I] and log $R'_{HK}$ values are reasonably well correlated, as expected 
if strong lithium absorption and strong chromospheric emission, as traced by Ca II 
emission, both indicate stellar youth.  The EW[H$\alpha$] and log $R'_{HK}$ values
are at best weakly correlated; stars with larger log $R'_{HK}$ values have, on average,
smaller EW[H$\alpha$] values, suggesting they are partially filled in by emission.  The
strong spectral type dependence of this temperature sensitive line likely masks much
of the underlying correlation which might be revealed by considering an
analogously defined quantity $R'_{H\alpha}$.  
The $v$sin$i$ values are well correlated,
in the mean, with log $R'_{HK}$ values.  This is expected since increased rotation is
believed to cause increased chromospheric activity, a characteristic common in young
stars.  Finally, in Figure \ref{fig_lithium_age} are shown the distributions of 
EW[H$\alpha$] 
and $v$sin$i$ versus EW[Li\,I].  Both the EW[H$\alpha$] and $v$sin$i$ values are 
correlated with EW[Li\,I], in the mean, though both show a large
scatter at all EW[Li\,I] values.

The above correlations corroborate the suggestion that these measured quantities
trace stellar age.  While the large scatter in these correlations inhibits assigning specific 
ages to individual stars, the ensemble distributions can help assess the age of the
observed sample.  Since the stellar temperature can also affect these quantities, 
we consider these distributions as a function of $B-V$ color, a proxy for stellar temperature.
The range of observed values at any $B-V$ color should therefore (mostly) be a 
consequence of the age spread in the sample; a comparison of the spread of values at 
different colors can serve as a check for age biases with respect to temperature.

\begin{figure}
\epsscale{1.0}
\plotone{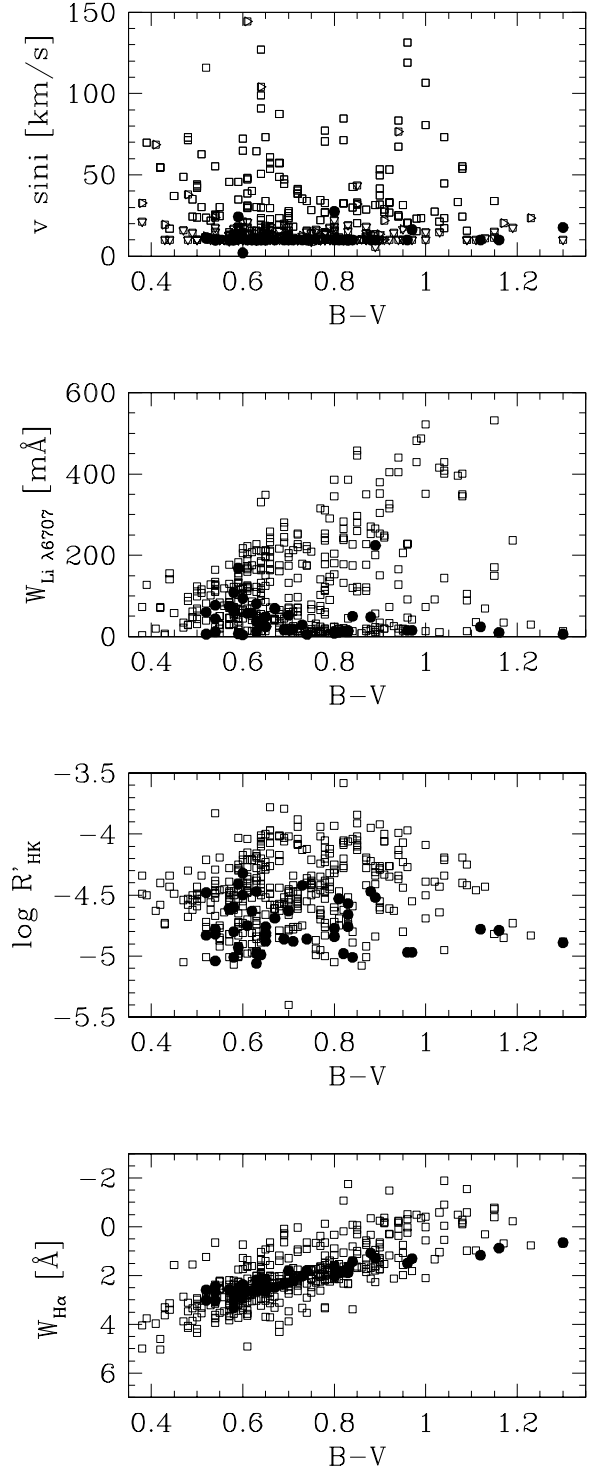}
\caption{Projected rotational velocity ($v$sin$i$), EW[Li\, $\lambda$6708], 
log $R'_{HK}$ and EW[H$\alpha$] versus $B-V$ color.  \textit{Filled circles} indicate 
dwarf standard stars.  For clarity, a few stars in the observed sample are bluer 
than (SAO 41857) or redder than (HIP 108467, HIP 90190, and  HIP 60783 ) 
the B-V range shown here, the values are nevertheless included in Table 1.  
Similarly, in the bottom 
panel four stars with strong H$\alpha$ emission (EW[H$\alpha$] $< -10$; Figure 
4) are not shown.
\label{fig_bmv_age} }
\end{figure}

In Figure \ref{fig_bmv_age} are shown the distributions of $v$sin$i$, 
EW[Li\,I $\lambda$6708], log $R'_{HK}$, and EW[H$\alpha$] versus $B-V$ 
color (as assembled in Section 4.4).  The observed dwarf standards are 
distinguished in order to illustrate the properties of old slowly rotating stars.
The distributions of $v$sin$i$ and log $R'_{HK}$ are independent of $B-V$
color, indicating a similar spread in rotation rate and chromospheric activity
over the range of spectral types observed.  The EW[Li\,I] values span a 
broad range at each $B-V$ color.  The larger EW[Li\,I] values at the red 
end is a temperature effect; for a specified lithium abundance,
the EW[Li\,I] increases with decreasing stellar temperature \citep{pm96}.  
The EW[H$\alpha$] values show the strongest $B-V$ trend, as 
expected for this temperature sensitive line; the EW[H$\alpha$] values 
decrease with increasing $B-V$ color.  There is also a larger scatter at
larger $B-V$ values.  The scatter above the dwarf loci can be interpreted as 
stars with some H$\alpha$ emission, whether this is partially or completely 
(e.g. EW[H$\alpha$] $< 0.0$) filling in the photospheric absorption line.  Four 
stars with strong H$\alpha$ emission (EW[H$\alpha$] $> -10$ \AA) are not 
plotted; their profiles are shown in Figure \ref{fig_halpha}.  Stars with some 
H$\alpha$ emission span the full range of $B-V$
colors, though there appear to be many more at redder colors ($B-V > 1.0$)
than bluer colors ($B-V < 0.6$).  However, since an EW is defined relative to 
the continuum flux, a unique EW[H$\alpha$] value corresponds to 
much less H$\alpha$ flux for cooler (red) stars than hotter (blue) stars.
Overall, these age indicators are consistent
with the observed sample having ages that span from a few million years
to a few billion years, as was intended.  Moreover, all spectral types appear
to span this range of ages; there is no strong age bias with spectral type
amongst the observed sample.

\subsection{Ca\,II H\&K Saturation in Young Stars}

While log $R'_{HK}$ values have been shown to correlate with stellar age 
\citep[e.g.][]{soderblom91}, this diagnostic
typically has not been used to estimate ages less than $\sim 100$ Myr.  This primarily stems 
from the lack of $R'_{HK}$ measurements for stars in young clusters of known age, from 
which the age relation could be calibrated.  Although the P60 $R'_{HK}$ measurements 
of young cluster members presented here may help establish this calibration, we illustrate
that accurate ages will be inhibited by the broad range in rotation rates for stars at a 
given young age and the saturation of flux at high rotation rates.

\begin{figure}
\epsscale{1.0}
\plotone{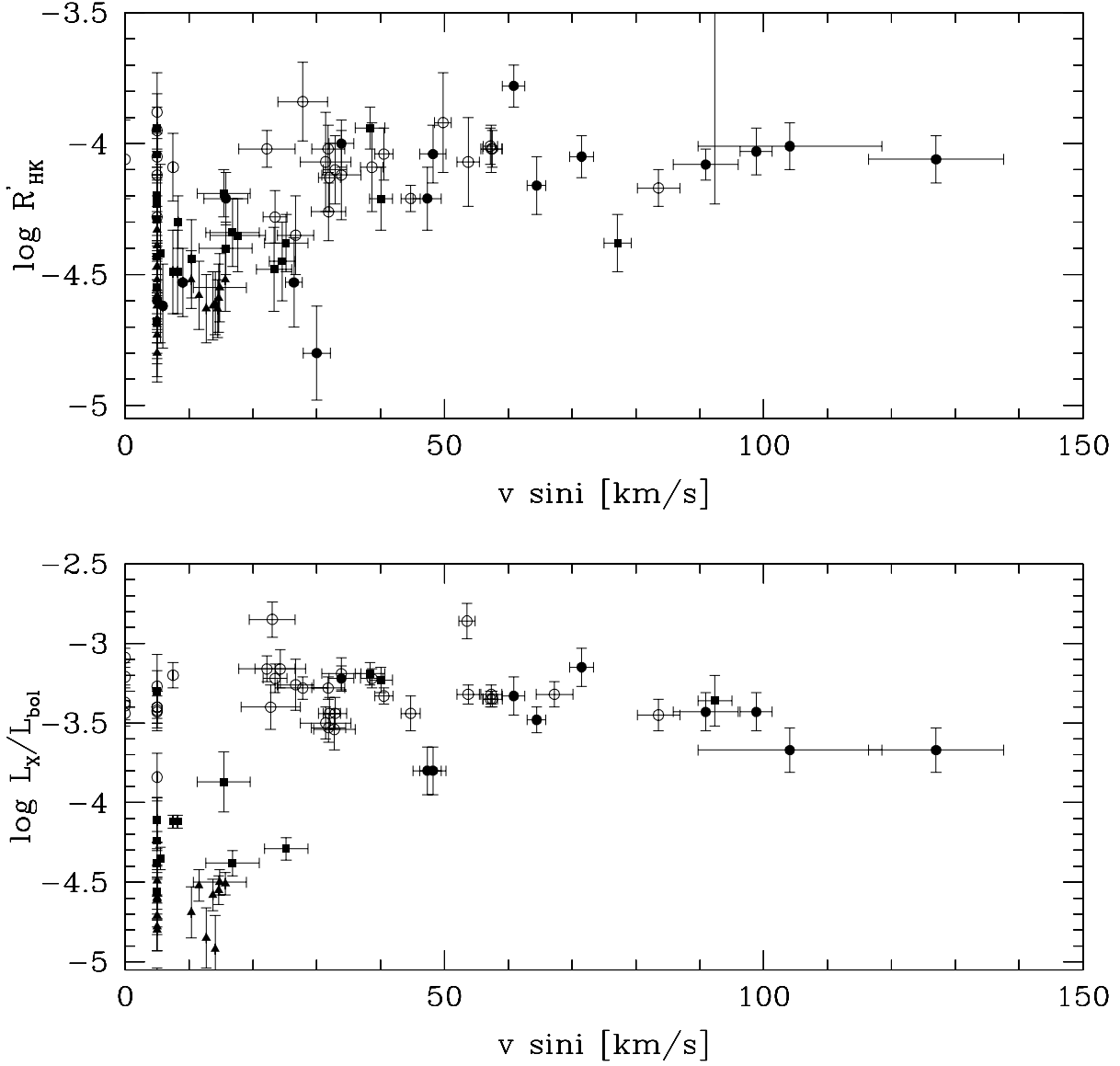}
\caption{$R'_{HK}$ and $L_X$/$L_{Bol}$ measurements of open cluster members 
versus $v$sin$i$.  \textit{Filled circles} are $\alpha$ Per members, \textit{squares}
are Pleiads, \textit{triangles} are Hyads and \textit{open circles} are T Tauri stars
in Upper Scorpius, Upper Centaurus Lupus, and R Corona Australis.  Stars with a
$v$sin$i$ upper limit are plotted with a $v$sin$i$ of 5 km s$^{-1}$ for clarity.
\label{fig_saturate} }
\end{figure}

In Figure \ref{fig_saturate} are shown the $R'_{HK}$ measurements versus projected
rotational velocity ($v$sin$i$) of stars in the open clusters represented in our sample, 
including $\alpha$ Persei,  
the Pleiades, the Hyades and collectively (post) T Tauri stars in Upper Scorpius, Upper Centaurus 
Lupus, and R Corona Australis.  As expected, the $R'_{HK}$ measurements are correlated
with $v$sin$i$; the most rapidly rotating stars have, on average, the largest $R'_{HK}$ 
values.  However, Figure \ref{fig_saturate} also indicates that once the $v$sin$i$ values
increase above $\sim 30$ km s$^{-1}$, the $R'_{HK}$ values stay roughly constant, 
implying no increase in the Ca\,II H\&K emission line flux toward larger rotation rates.
A similar ``saturation" phenomenon has been seen before based on observations of other 
emission lines originating in stellar chromospheres (OI\,1304, C\,II 1335), as well as X-rays 
originating in stellar corona \citep[e.g.][]{vilhu84}.  It is not yet known whether the cause
for flux saturation is internal (e.g. dynamo) or external (e.g. active regions) to the star 
\citep[see][]{gudel04}.  Saturation of the Ca\,II H\&K fluxes, as measured by the $R'_{HK}$ 
diagnostic, has not been demonstrated before.

While saturation of the Ca\,II H\&K flux seems consistent with other chromospheric 
and coronal observations of rapidly rotating stars, we considered the possibility that 
this saturation is simply a consequence of the method used to determine
the $S$ values, which are used to calculate $R'_{HK}$.  Specifically, since the 
prescription outlined in Section 4.4 involves measuring the Ca\,II K
and H emission with a triangular passband of specified width, apparent saturation 
could be caused by emission line flux being rotationally broadened outside of
this passband, and thus not included in the measurement.  To test this, we remeasured S values for a
set of stars that were artificially rotationally broadened from 10 to 150 km s$^{-1}$; we assume
the core emission is rotationally broadened by the same amount as the photosphere.  This
showed that the technique used to measure the $S$ values begins to miss core emission
flux above $v$sin$i$ of $\sim 100$ km s$^{-1}$.  Since the saturation observed for
$R'_{HK}$ occurs well below this value, we interpret the saturation as a real effect.

To further demonstrate the case for saturation, in Figure \ref{fig_saturate} we also
illustrate the ratio of X-ray flux to bolometric flux ($L_X/L_{bol}$) for the stars in these
clusters which have the available data.  X-ray fluxes are from ROSAT observations 
and bolometric
fluxes are determined from the atmospheric model which best fits the stellar energy 
distribution (see Mamajek et al. 2007 for a complete description of the $L_X/L_{bol}$ 
derivation).  A saturation trend is identified for coronal X-rays similar to that for chromospheric
Ca\,II H\&K core emission.  The most rapidly
rotating stars are the most fractionally X-ray bright, but $L_X/L_{bol}$ remains roughly
constant above $v$sin$i$ values of $\sim 30$ km s$^{-1}$.  This phenomenon has been
demonstrated previously by many authors.  What we have newly demonstrated here is 
that the same young cluster
stars that have saturated X-ray emission also have saturated chromospheric Ca\,II H\&K 
emission.

\acknowledgements
We are grateful to the observing assistants formerly present at the Palomar 60-inch, 
Karl Dunscombe 
and Skip Staples, for their help and patience in acquiring this large data set. 
We thank E. Mamajek and C. Slesnick for assisting with the observations and 
Jonathan Foster, Francesca Colonnese, and Michael Inadomi for their assistance with 
data reduction and analysis.  We appreciate greatly the scattered light analysis
provided by J. Valenti for some of our spectra.
This research has made use of the SIMBAD and 2MASS 
databases.  We acknowledge support for these observations from NASA contract 
1224566 administered through JPL.  
This work has benefited from the ability to organize our data within the FEPS 
database created by John Carpenter.

\clearpage

\begin{deluxetable}{lllllllcccllccccc}	
\rotate
\tabletypesize{\tiny}	
\tablecaption{Spectroscopic Properties of Observed Single(-Lined) Stars\label{tab_single} }	
\tablewidth{0pt}	
\tablehead{ \colhead{HD} 	
& \colhead{HIP}	
& \colhead{other}	
& \colhead{Grp}	
& \colhead{RA(2000)}	
& \colhead{DEC(2000)}	
& \colhead{Epoch}	
& \colhead{SNR}	
& \colhead{EW[Li\,I]}	
& \colhead{EW[H$\alpha$]}
& \colhead{RV}	
& \colhead{vsini}	
& \colhead{SNR}	
& \colhead{}	
& \colhead{Spectral}	
& \colhead{}	
& \colhead{} \\	
\colhead{} 	
& \colhead{}	
& \colhead{}	
& \colhead{}	
& \colhead{}	
& \colhead{}	
& \colhead{}	
& \colhead{$at Li$}	
& \colhead{(\AA)}	
& \colhead{(\AA)}	
& \colhead{(km s$^{-1}$)}	
& \colhead{(km s$^{-1}$)}	
& \colhead{$at HK$}	
& \colhead{$S_{MW}^a$} 
& \colhead{Type$^b$}
& \colhead{B-V$^c$}	
& \colhead{R'$_{HK}^{a,d}$} }
\startdata
224873	& 110	& RX J0001.4+3936	& Field & 00 01 23.66 & +39 36 38.1 & 2001 Dec 2 	 & 105 & $0.032$ & 1.74 & $-5.6\pm1.4$ & $<16.3$		& 50 & 0.365 0.068 & K0 & 0.76t & -4.49 0.11 \\
105	& 490	        & RX J0005.9-4145	& Field & 00 05 52.55 & -41 45 10.9 & 2001 Dec 2 	 & 217 & $0.156$ & 3.06 & $1.6\pm1.2$ & $<18.2$		& 16 & 0.675e 0.118 & G0V & 0.59t & -4.03e 0.09 \\
377	& 682 & 1RXS J000826.8+063712	& Field & 00 08 25.74 & +06 37 00.5 & 2001 Jul 27 	 & 348 & $0.122$ & 2.34 & $-3.8\pm2.2$ & $<12.0$		& 135 & 0.348 0.12 & G2V & 0.63t & -4.41 0.20 \\
691	& 919	        & V344 And		& Field & 00 11 22.40 & +30 26 58.4 & 2002 Oct 29  & 256 & $0.122$ & 2.20 & $-2.4\pm0.8$ & $<10.$		& 84 &  0.388 0.072 & K0V & 0.78t & -4.47 0.10 \\
	&	&		&				&	   &		    				 & 2002 Sep 18 & 174 & $0.115$ & 2.14 & $-3.0\pm0.8$ & $<13.7$		& 84 &0.391 0.073 & " & 0.78t & -4.47 0.10 \\
984	& 1134 & 1RXS J001410.1-071200	& Field & 00 14 10.25 & -07 11 57.0 & 2001 Jul 27 	 & 315 & $0.099$ & 2.99 & $-3.2\pm2.2$ & $42.1\pm1.7$& 131 & 0.296 0.104 & F7V & 0.50t & -4.43 0.22 \\
1326A	& 1475	& GJ 15 A			& Mstar & 00 18 22.57 & +44 01 22.2 & 2002 Oct 30  & 198 & $<0.01$ & 0.51 & $12.0\pm0.9$ & $<20.5$		& 38 & 0.568 0.101 & M2V & 1.48t & \nodata \\
1326B	&\nodata  & GJ 15 B			& Mstar & 00 18 25.50 & +44 01 37.6 & 2002 Oct 30  & 87 &  $<0.02$ & 0.17 & $10.7\pm0.9$ & $<18.9$		& 14 & 1.083 0.181 & M5V & 1.61s & \nodata \\
3765	& 3206	        & GJ 28			& Field & 00 40 49.29 & +40 11 13.3 & 2002 Sep 17 & 259 & $0.015$ & 1.31 & $-63.8\pm0.7$ & $16.3\pm5.1$& 65 & 0.218 0.043 & K2V & 0.97t & -4.97c 0.10 \\
\nodata	&\nodata  & QT And			& Field & 00 41 17.32 & +34 25 16.8 & 2002 Oct 29  & 160 & $0.128$ & -0.17 & $6.0\pm0.8$ & $24.4\pm2.2$& 40 & 1.617 0.259 & G & 0.94t & -3.99c 0.11 \\
		&		&				&	    &			   &			 & 2002 Sep 18  & 83 & $0.166$ & -0.23 & $5.1\pm0.7$ & $25.1\pm2.2$& 27 &  0.705 0.123 & " & 0.94t & -4.37c 0.11 \\
6434	& 5054	        & GJ 9037		& Field & 01 04 40.15 & -39 29 17.4 & 2001 Dec 2    & 231 & $<0.004$ & 2.37 & $21.9\pm1.2$ & $<10.$			& 28	&0.396 0.074 & G2/3V & 0.60t & -4.32 0.11 \\
6963	& 5521	        & SAO 36995		& Field & 01 10 41.92 & +42 55 54.7 & 2002 Sep 17 & 206 & $<0.003$ & 2.05 & $-31.7\pm0.7$ & $<12.58$	& 77	&  0.226 0.045 & G7V & 0.73t & -4.75 0.14 \\
7661	& 5938	        & SAO 147702		& Field & 01 16 24.18 & -12 05 49.3 & 2003 Feb 8 	 & 185 & $0.075$ & 2.04 & $5.2\pm0.4$ & $<10.$		& 33	& 0.479 0.087 & K0V & 0.77t & -4.35 0.10 \\
\nodata	& 6276	& SAO 147747		& Field & 01 20 32.26 & -11 28 03.6 & 2003 Feb 8 	 & 178 & $0.153$ & 1.99 & $8.3\pm0.4$ & $<10.$		& 29	& 0.619 0.110 & G0 & 0.80t & -4.25 0.09 \\
8467	& 6575	        & SAO 54666		& Field & 01 24 27.98 & +39 03 43.6 & 2002 Sep 17 & 172 & $<0.011$ & 1.64 & $15.0\pm0.7$ & $<10.$		& 60	& 0.278 0.054 & G5 & 0.77t & -4.65 0.12 \\
8941	& 6869	        & SAO 92453		& Field & 01 28 24.36 & +17 04 45.2 & 2001 Jul 29 	 & 421 & $0.006$ & 2.59 & $9.3\pm2.2$ & $<10.96$		& 181& 0.175 0.035 & F8IV-V & 0.52t & -4.83 0.18 \\
	&			&				&	     &			   &			 & 2002 Sep 18 & 233 & $<0.004$ & 2.69 & $8.29\pm0.70$ & $<12.3$	& 123& 0.190 0.069 & F8IV-V & 0.52t & -4.76 0.31 \\
8907	& 6878	        & IRAS 01256+4200	& Field & 01 28 34.35 & +42 16 03.8 & 2001 Jul 26 	 & 266 & $0.048$ & 1.55 & $6.3\pm2.2$ & $<14.34$		& 142 & 0.263 0.093 & F8 & 0.49t & -4.50 0.24 \\
9472	& 7244	        & SAO 74789		& Field & 01 33 19.03 & +23 58 32.1 & 2002 Sep 18 & 190 & $0.055$ & 2.28 & $10.9\pm0.7$ & $19.33\pm4.22$&	83 &  0.277 0.053 & G0 & 0.68t & -4.59 0.13 \\
\nodata	&\nodata  & RX J0137.6+1835	& Field & 01 37 39.41 & +18 35 33.2 & 2001 Jul 28 	 & 194 & $0.416$ & -0.59 & $1.8\pm2.2$ & $21.25\pm3.26$&	26 & 0.512e 0.171 & K3Ve & 1.03t & -4.64e,c 0.23 \\
10780	& 8362	& IRAS 01441+6336	& Field & 01 47 44.88 & +63 51 09.1 & 2001 Jul 28 	 & 203 & $0.012$ & 1.66 & $1.4\pm2.2$ & $<10.$		&55	& 0.237 0.085 & K0V & 0.80t & -4.77 0.22 \\
11850	& 9073	& SAO 75038		& Field & 01 56 47.28 & +23 03 04.1 & 2002 Oct 29 & 176 & $0.018$ & 2.32 & $3.3\pm0.8$ & $<12.08$		& 54	& 0.287 0.055 & G5 & 0.69t & -4.57 0.12 \\
		&		&				&	    &			   &			 & 2002 Sep 18 & 138 & $0.015$ & 2.27 & $1.6\pm0.7$ & $<12.38$		& 76	& 0.292 0.056 & " & 0.69t & -4.56 0.12 \\
\enddata	
\tablenotetext{a}{Stars with especially uncertain S values are marked with an "e".}
\tablenotetext{b}{Stars which appear low surface gravity are marked with an asterisk.}	
\tablenotetext{c}{B-V colors from the Tycho catalogue are marked with a "t" while those from the stellar spectral type are marked with an "s".}
\tablenotetext{d}{Stars with B-V colors outside the range over which R'$_{HK}$ values are 
calibrated are marked with a "c".   }
\end{deluxetable}

\begin{deluxetable}{lllllllcccc}
\tabletypesize{\tiny}	
\tablecaption{Observed Double-Lined Binary Stars \label{tab_binary} }	
\tablewidth{0pt}	
\tablehead{ \colhead{}
& \colhead{}	
& \colhead{}
& \colhead{}
& \colhead{}	
& \colhead{}	
& \colhead{}	
& \colhead{SNR}
& \colhead{EW[Li\.I]}	
& \colhead{EW[Ca\,I]}	
& \colhead{EW[H$\alpha$} \\
\colhead{HD}
& \colhead{HIP}	
& \colhead{other}	
& \colhead{Assoc}
& \colhead{RA(2000)}
& \colhead{DEC(2000)}	
& \colhead{Epoch}	
& \colhead{$at Li$}	
& \colhead{(m\AA)}	
& \colhead{(m\AA)}	
& \colhead{(\AA)} }
\startdata	
\nodata & \nodata & Cl~Melotte~20~848	& aPer & 03 29 26.24 & +48 12 11.74	&2002 Oct 31	& 149 & 45/69 & 61/85 & 1.65/0.95 \\	
78899   & 45187   & GJ 3539 			& Field & 09 12 28.24 & +49 12 23.4		&2003 Feb 9	& 260 & 11/50 & 91/121 & 0.83/0.72\\	
\nodata & \nodata & SAO 178272		& Field& 09 59 08.42 & -22 39 34.57	&2002 Feb 4	& 107 & 87/162 &105/153& em.\\	
"		& "		& "				& "		& "		& "				&2001 Dec 2	& 134 &\nodata &\nodata & em.\\
"		& "		& "				& "		& "		& "				&2002 Apr 19	& 51 &103/204&174/233 & -0.1/-0.2 \\
\nodata & \nodata & TYC 7310 503 1	& Field & 14 58 37.69 & -35 40 30.27	&2003 Jun 5	& 82 & 121/222& 99/180 & em.\\
\nodata & \nodata & TYC 7305 380 1	& Field & 14 50 25.82 & -35 06 48.66	&2002 Feb 3	& 85 & 132/250& 103/66 &\nodata \\
"		& "		& "				& "		& "		& "				&2003 Jun 5	& 79 & 307/60   & 120/54 &\nodata \\
140374 & 77081 & TYC 7331 1235 1	& Field & 15 44 21.06 & -33 18 54.97	&2003 Jun 4	&  145 & 112/116 & 61/77  & -0.96/-1.17\\
142229 & 77810   & SAO 121238		& Field & 15 53 20.01 & +04 15 11.5		&2002 Feb 2	& 119& 56/121  & 29/79  &\nodata\\
218738 & 114379 & KZ And 			& Field & 23 09 57.34 & +47 57 30.0		&2003 Jun 6	& 73 &  69/107   & 108/165 &\nodata \\
\enddata
\end{deluxetable}

\end{document}